\documentclass[a4paper,11pt]{article}
 
\usepackage{fullpage}
\usepackage{upgreek}
 
\usepackage{authblk}
\usepackage{amsmath,amssymb, amsthm,mathrsfs}
\usepackage{graphicx}
\graphicspath{{./fig/}}
 
\usepackage{verbatim}
\usepackage{complexity}
\usepackage{color}
\usepackage{algorithm}
\usepackage{algorithmic}
\usepackage{url}
\usepackage{wasysym}
\usepackage{subcaption,xcolor}

\usepackage{mathdots}
\usepackage{tikz}
\usepackage[normalem]{ulem}
\usetikzlibrary{arrows.meta,fit,calc,backgrounds,decorations.pathmorphing,positioning}
\usepackage[textsize=small]{todonotes}

\newtheorem{theorem}{Theorem}

\newtheorem{claim}[theorem]{Claim}
\newtheorem{question}[theorem]{Question}
\newtheorem{corollary}[theorem]{Corollary}
\newtheorem{lemma}[theorem]{Lemma}

\newcommand{\qedclaim}{\hfill $\diamond$ \medskip}
\newenvironment{proofclaim}{\noindent{\it Proof of Claim.}}{\qedclaim}

\DeclareMathOperator{\h}{\mathcal{H}}
\DeclareMathOperator{\I}{\textrm{I}}

\DeclareMathOperator{\hn}{hn}

\DeclareMathOperator{\itn}{in}
 
\DeclareMathOperator{\smc}{sm}
\DeclareMathOperator{\mc}{m}

\usepackage{esvect}

\DeclareMathOperator{\ohn}{\vv{\hn}}
\DeclareMathOperator{\oin}{\vv{\itn}}
 
\DeclareMathOperator{\oinm}{\oin_{\mc}}
\DeclareMathOperator{\ohnm}{\ohn_{\mc}}

\DeclareMathOperator{\ohnsm}{\ohn_{\smc}}
 
\DeclareMathOperator{\oi}{\vv{\I}}
\DeclareMathOperator{\oifsm}{\oi_{\smc}}
\DeclareMathOperator{\oifm}{\oi_{\mc}}

\DeclareMathOperator{\hull}{H}

\DeclareMathOperator{\ohull}{\vv{\hull}}
\DeclareMathOperator{\ohullsm}{\ohull_{\smc}}
\DeclareMathOperator{\ohullm}{\ohull_{\mc}}

\usepackage{pbox}
\usepackage{hyperref}
\usepackage{tabularx}
\usepackage{nameref}
\usepackage[skins,listings]{tcolorbox}

\newtcolorbox{ProblemBox}[2][]{
	colframe=black!20!white,
	coltitle=black,
	arc=1.5mm,
	boxrule=.15mm,
	colbacktitle=white,
	colback=white,
	enhanced,
	adjusted title=flush left,
	attach boxed title to top left={yshift=-3mm,xshift=3mm},
	boxed title style={colframe=white,righttitle=-3mm,lefttitle=-3mm},
	title=#2,#1
}

\author[1]{Júlio Araújo}
\author[1]{Ana Karolinna Maia}
\author[3]{Thiago Marcilon}
\author[2]{Nicolas Nisse}
\author[1]{Rudini Sampaio}
 
\affil[1]{ParGO Research Group, Universidade Federal do Cear\'a, Fortaleza, Brazil}
\affil[2]{Universit\'e C\^ote d'Azur, CNRS, Inria, I3S, Sophia Antipolis, France}
\affil[3]{Universidade Federal do Cariri, Juazeiro do Norte, Brazil}

\title{The Monophonic Directed Convexities\thanks{Research supported by the CAPES-Cofecub project Ma 1004/23, CAPES Finance Code 001, by the French government, through the EUR DS4H Investments in the Future project managed by the National Research Agency (ANR) with the reference number ANR-17-EURE-0004, and by Conselho Nacional de Desenvolvimento Científico e Tecnológico (CNPq) 313558/2025-6. J\'ulio Ara\'ujo is funded by CNPq 308939/2025-5 and 404479/2023-5.}}
 
\begin{document}
\maketitle
 
\begin{abstract}
Graph Convexity is a topic with many papers in the literature, especially on the $P_3$, geodesic, and monophonic convexities in undirected graphs. Regarding directed graphs, Erd{\H o}s et al. in 1972 and Chartrand and Zhang in 2000 obtained the first results on the $P_3$ and geodesic convexities for directed graphs, respectively.
In this paper, we start the investigation of monophonic convexity on directed graphs.
We define two convexities, the monophonic and the strong monophonic directed convexities, related to the minimal and the induced directed paths, respectively.
We show that deciding whether a set is convex is polynomial-time solvable in the monophonic convexity, but $\coNP$-complete in the strong monophonic convexity.
For both convexities, we prove that the hull number is $\NP$-complete even in DAGs.
Regarding the monophonic convexity, we prove that the hull number is polynomial time solvable in tournaments, but the interval number is $\NP$-complete in tournaments and also in bipartite oriented graphs.
\end{abstract}
 
\section{Introduction}
\label{sec:intro}
 
Convexity spaces are a classic topic studied in different branches of mathematics.
The research on convexities applied to graphs is a more recent topic, dating back about 50 years. Classical convexity definitions and results motivated the definitions of some graph parameters (Carathéodory number~\cite{barbosa-2012}, Helly number~\cite{DOURADO2017134}, Radon number~\cite{Delire1984}, rank~\cite{Kante+2017}, hull number~\cite{THNOAG}, and interval number~\cite{jayme18}).
The complexity aspects related to the determination of the values of those parameters have been the central issue of several recent works in this field.
For more information, we recommend the recent book \cite{araujo-livro25}.
 
A \emph{graph convexity} $\mathcal{C}$ on a finite graph $G$ can be defined by an interval function $\I_\mathcal{C}:2^{V(G)}\to 2^{V(G)}$ such that $\I_\mathcal{C}(\emptyset)=\emptyset$, $S\subseteq\I_\mathcal{C}(S)$, and $\I_\mathcal{C}(S)\subseteq\I_\mathcal{C}(S')$ for every $S\subseteq S'\subseteq V(G)$. We say that a subset $S\subseteq V(G)$ is \emph{convex} in the convexity $\mathcal{C}$ if $\I_\mathcal{C}(S)=S$.
The \emph{convex hull} $\hull_\mathcal{C}(S)$ of a subset $S\subseteq V(G)$ is the minimum convex set that contains $S$.
We say that a subset $S\subseteq V(G)$ is a \emph{hull set} (resp. \emph{interval set}) in the convexity $\mathcal{C}$ if $\hull_\mathcal{C}(S)=V(G)$ (resp. $\I_\mathcal{C}(S)=V(G)$). Note that in every $n$-vertex graph, $\hull_\mathcal{C}(S)=\I^n_\mathcal{C}(S)$, that is, the convex hull can be obtained through at most $n$ applications of the interval function.
The hull number $\hn_\mathcal{C}(G)$ is the size of a minimum hull set of $G$, and the interval number $\itn_\mathcal{C}(G)$ is the size of a minimum interval set of $G$.
These definitions work for directed and undirected graphs.
 
The most investigated graph convexities are the so called \emph{path convexities}. Given a family $\mathcal{P}$ of paths in a graph, the convexity induced by $\mathcal{P}$ is such that $\I_\mathcal{C}(S)$ contains $S$ and every vertex in a path of $\mathcal{P}$ between two vertices of $S$.
The vast majority of works on graph convexity focus on path convexities in undirected graphs, such as the geodesic, monophonic, $P_3$, and $P_3^*$ convexities.
The geodesic convexity is the convexity related to shortest paths \cite{Harary1}; the monophonic convexity is related to minimal (i.e., induced) paths \cite{jamison1982}; the $P_3$ convexity is related to paths with 3 vertices \cite{CDS09}; and the $P_3^*$ convexity is related to minimal paths with 3 vertices \cite{jayme18}.
 
Erd{\H o}s et al.~\cite{SROST} obtained one of the first results in graph
convexity in 1972, which focused on the $P_3$ convexity in tournaments: directed graphs in which there is exactly one arc between every pair of vertices. See also~\cite{EITIST,PFALTZ1,CIT} for papers from the 1970's.
Despite this fact, graph convexity in directed graphs has received much less attention than
in undirected graphs.
The $P_3$ convexity in tournaments was revisited in the 1990's~\cite{HW1996} and
later extended to multipartite tournaments~\cite{PWW2006,PARKER2008641,PWW2009}.
Chartrand and Zhang~\cite{TGNOAOG} and Chartrand et al.~\cite{CFZ2002,THNOAOG}
obtained  the first results on the geodesic directed convexity in the early 2000's,
introducing the geodesic number and the hull number of oriented graphs, which were then
investigated for orientations of graphs~\cite{OCGAHNIG,THAGNOOOG,TGNOGAD,SCSgrids2017}.
In the same period, twin dominating sets~\cite{Arumugam2013,Dankelman2003}, which are
exactly the interval sets of the $P_3$ convexity in directed graphs, were introduced, and convex sets
with respect to all directed paths in acyclic digraphs were studied~\cite{BGG2009,BGGJRSSY2009}.
More recently, algorithmic aspects of the geodesic convexity in oriented graphs were
investigated~\cite{AA2021,AMMP2024arXiv,FLMGTP2026}, and the oriented graphs whose
$P_3$ and $P_3^*$ convexities are convex geometries were characterized~\cite{PAGDT2026}.
 
\subsection{Our results}

In this paper, we start the investigation of the monophonic convexity on directed graphs.
Let $D$ be a directed graph.
For simplicity, an arc $(u,v)$ of $D$ is sometimes denoted by $uv$. Similarly, a directed path with vertices $v_1,\ldots,v_\ell$ (in this order) is denoted by $(v_1,\ldots,v_\ell)$ or, to emphasize its direction, by $v_1\to\cdots\to v_\ell$.
We say that $D$ is an \emph{oriented graph} if, for every arc $v_iv_j$, we have that $v_jv_i$ is not an arc.
A directed path $(v_1,\ldots,v_{\ell})$ is \emph{minimal} if there is no arc $v_iv_j$ with $i<j-1$.

Unlike in undirected graphs, a minimal path and an induced path do not have the same meaning in directed graphs.
According to Bang-Jensen and Gutin \cite{BG2008}, an induced path in a directed graph must also be induced in the corresponding underlying undirected graph.
For example, a path $v_1\to v_2\to v_3$ is minimal if $v_1v_3$ is not an arc ($v_3v_1$ can be an arc), but it is induced if $v_1v_3$ and $v_3v_1$ are not arcs.
Therefore, it is possible to define two convexities in directed graphs: the \emph{monophonic directed convexity} (related to minimal directed paths) and the \emph{strong monophonic directed convexity} (related to induced directed paths), denoted by $\vv{m}$ and $\vv{sm}$, respectively.
 
For simplicity, we prefer not to use arrows in subscripts. That is, given a directed graph $D$ and a set $S$ of vertices of $D$, we will use the notations $\oinm(D)$, $\ohnm(D)$, $\oifm(S)$ and $\ohullm(S)$, instead of  $\itn_{\vv{m}}(D)$, $\hn_{\vv{m}}(D)$, $\I_{\vv{m}}(S)$ and $\hull_{\vv{m}}(S)$, which are well defined.
For example, the \emph{monophonic interval} $\oifm(S)$ (resp.\ the \emph{strong monophonic interval} $\oifsm(S)$) consists of $S$ and every vertex of a minimal (resp.\ induced) directed path between two vertices of $S$.

We first show that deciding whether a set is convex is polynomial-time solvable in the monophonic directed convexity, but $\coNP$-complete in the strong monophonic directed convexity.
We prove that the hull number is $\NP$-complete even in DAGs for both convexities: monophonic and strong monophonic.
Regarding the monophonic convexity, we prove that the hull number is polynomial time solvable in tournaments, but the interval number is $\NP$-complete in tournaments and also in bipartite oriented graphs.
 
\section{Recognizing convex sets}\label{sec-NP}
 
In this section, we show that deciding whether a set is convex can be done in polynomial time in the monophonic convexity, but it is $\coNP$-complete in the strong monophonic convexity.
 
We first show the result of the monophonic convexity.

\begin{theorem}\label{thm:hull-poly}
Let $D$ be a directed graph on $n$ vertices.
Given a set $S$ of vertices of $D$, the monophonic convex hull $\ohullm(S)$ can be obtained in time $\mathcal{O}(n^4)$.
\end{theorem}

\begin{proof}
Let $V=V(D)$. One should notice that if a set $X$ of vertices is not convex in the monophonic convexity, then there exists a minimal path whose endpoints lie in $X$ and whose internal vertices lie in $V\setminus X$. Moreover, for any two distinct vertices $a,b\in X$ such that $ab\notin A(D)$, a shortest path from $a$ to $b$ in $D[V\setminus (X \setminus \{a,b\})]$ is minimal, since an arc $v_iv_j$ with $i<j-1$ would give a shorter path. Therefore, starting with $X=S$, it suffices to iteratively look for a shortest path in $D[V\setminus (X \setminus \{a,b\})]$ linking $a$ to $b$, for any $a,b\in X$ such that $ab\notin A(D)$, and add its vertices to $X$, until no such path exists. At the end, $X=\ohullm(S)$. Each successful search adds at least one vertex to $X$, and a failed search for a pair $a,b$ also fails later, since $X$ only grows. Then there are $\mathcal{O}(n^2)$ searches by BFS, each one in time $\mathcal{O}(n^2)$, which gives time $\mathcal{O}(n^4)$.
\end{proof}

By Theorem~\ref{thm:hull-poly}, deciding whether $\ohnm(D)\leq k$ is in $\NP$. The proof of this theorem relies on the fact that a shortest directed path is minimal. This is not the case for induced paths, since a shortest directed path may have arcs in the opposite direction.
In fact, Theorem~\ref{thm:hull-poly} cannot be extended to the strong monophonic convexity, unless $\P=\NP$.

\begin{theorem}\label{thm:sm-conp}
Let $D$ be a directed graph.
Then, given a set $S$ of vertices of $D$, deciding whether $S$ is convex in the strong monophonic convexity is $\coNP$-complete. Consequently, there is no polynomial-time algorithm that computes the strong monophonic convex hull $\ohullsm(S)$, unless $\P = \NP$.
\end{theorem}
 
\begin{proof}
The problem is in $\coNP$: if $S$ is not convex, a certificate is an induced directed path between two vertices of $S$ containing a vertex not in $S$, which can be checked in polynomial time.
In \cite[Lemma~7]{BHT2012}, it was proved that it is $\NP$-complete to decide whether an oriented graph contains an induced directed path from $a$ to $b$, for prescribed vertices $a$ and $b$. Adding new vertices $s$ and $t$ and new arcs $(s,a)$ and $(b,t)$, we have that $\{s,t\}$ is not convex in the strong monophonic convexity if and only if there is an induced directed path from $a$ to $b$, since $s$ (resp.\ $t$) is adjacent only to $a$ (resp.\ $b$). Hence, deciding whether $\{s,t\}$ is convex is $\coNP$-complete.
\end{proof}
 
Despite that, it is possible to show that deciding whether $\ohnsm(D)\leq k$ is in $\NP$: a certificate is a hull set $S$ with $|S|\leq k$ together with a sequence of at most $n$ induced directed paths, each one between two vertices of $S$ or of previous paths, which together cover $V(D)$.

\section{Monophonic hull is $\NP$-hard in DAGs}
 
In this section, we prove that the hull number is $\NP$-hard in both monophonic and strong monophonic convexities.
Notice that these convexities are equivalent in DAGs (directed acyclic graphs), since minimal paths and induced paths are equivalent in DAGs.
 
\begin{theorem}
Given a directed graph $D$ and an integer $k$, deciding if the hull number of $D$ is at most $k$ is $\NP$-complete for both monophonic and strong monophonic convexities, even in DAGs.
\end{theorem}
 
\begin{proof}
The problem is in $\NP$ by Theorem~\ref{thm:hull-poly}, since both convexities coincide in DAGs.
We obtain a reduction from Restricted-3SAT, which is known to be $\NP$-complete \cite{dahlhaus94}, inspired by the reduction of the $P_3$ hull problem in \cite{jayme18}. In Restricted-3SAT, each clause has at most three literals, each variable appears in at most three clauses and every literal of the form $\overline{x}$, where $x$ is a variable, appears in exactly one clause. Let $\Phi$ be a formula of Restricted-3SAT with variables $x_1,\ldots,x_n$ and clauses $C_1,\ldots,C_m$. Let $\ell$ be the number of variables which appear in exactly 3 clauses.
 
We construct a DAG $D$ from $\Phi$ such that $\Phi$ is satisfiable if and only if $D$ has monophonic hull set of size $14n+m+\ell$.
The variable gadget related to variable $x_i$ is shown in Figure \ref{fig-hull-var}. The clause gadgets and the relation with the variable gadgets are shown in Figure \ref{fig-hull-clause}.

Let $S$ be the set of all sinks and sources of $D$.
The construction of $D$ is in such a way that there is an arc from every source of $D$ to every sink of $D$.
The vertices of $S$ will be represented in black and the arcs between black vertices are omitted.
Non-black vertices are called white vertices.
 
Regarding the clause gadgets, add for every clause $C_p$ the vertices $c_p$ and $z_p$ (which is a source), and the arc $(z_p,c_p)$.
If the literal $x_i$ (positive) is contained in two clauses $C_p$ and $C_q$, add the vertices $\alpha_i$, $\beta_i$ and $\gamma_i$ and the arcs shown in Figure \ref{fig-hull-clause}.
If a literal $x_i$ or $\overline{x_i}$ (say $\overline{x_i}$) is contained in only one clause $C_r$, only add the arc $(c_r,\overline{x_i})$, as shown in Figure \ref{fig-hull-clause}.
Notice that the vertices $\alpha_i$, $\beta_i$ and $\gamma_i$ only exist if the variable $x_i$ appears in exactly 3 clauses. Then there are $\ell$ vertices like $\gamma_i$.

\begin{figure}[ht]\centering
\begin{tikzpicture}[arc/.style={->, thick}]
\tikzstyle{v1}=[draw,circle,fill=black!10,minimum size=15pt,inner sep=1pt]
\tikzstyle{v2}=[draw,circle,fill=black!100,minimum size=8pt,inner sep=0pt]
 
\node[v1] (x) at (0,5.46) {$x_i$};
\node[v1] (a) at(-1,3.73) {$a_i$};
\node[v1] (b) at (1,3.73) {$b_i$};
\node[v1] (c) at (0,2) {$c_i$};
\node[v1] (d) at (0,0) {$d_i$};
\node[v1] (e) at (2,2) {$e_i$};
\node[v1] (f) at (2,0) {$f_i$};
 
\node[v1] (g) at (4,0) {$g_i$};
 
\node[v1] (x2) at (8,5.46) {$\overline{x_i}$};
\node[v1] (a2) at (9,3.73) {$\overline{a_i}$};
\node[v1] (b2) at (7,3.73) {$\overline{b_i}$};
\node[v1] (c2) at (8,2) {$\overline{c_i}$};
\node[v1] (d2) at (8,0) {$\overline{d_i}$};
\node[v1] (e2) at (6,2) {$\overline{e_i}$};
\node[v1] (f2) at (6,0) {$\overline{f_i}$};
 
\node[v2] (s0) at (2,4.4) {}; \node at (2.4,4.3) {$s_i$};
\node[v2] (s1) at (-1,5) {};
\node[v2] (s2) at (1,2.58) {};
\node[v2] (s3) at (1,1.5) {};
\node[v2] (s4) at (1,0.5) {};
\node[v2] (s5) at (1,-.5) {};
 
\node[v2] (s6) at (4,1) {};
 
\node[v2] (s0b) at (6,4.4) {};\node at (5.6,4.3) {$\overline{s_i}$};
\node[v2] (s1b) at (9,5) {};
\node[v2] (s2b) at (7,2.58) {};
\node[v2] (s3b) at (7,1.5) {};
\node[v2] (s4b) at (7,0.5) {};
\node[v2] (s5b) at (7,-.5) {};
 
\draw[arc] (s1) edge (x) edge (a);
\draw[arc] (s2) edge (b) edge (c) edge (e);
\draw[arc] (s4) edge (d) edge (f);
\draw[arc] (s6) edge (g);
\draw[arc] (s1b) edge (x2) edge (a2);
\draw[arc] (s2b)edge(b2)edge(c2)edge(e2);
\draw[arc] (s4b) edge (d2) edge (f2);
 
\draw[arc] (s2) to[out=150,in=-90] (x);
\draw[arc] (s2b)to[out= 30,in=-90] (x2);
 
\draw[arc] (x) to (s0);
\draw[arc] (a) edge (x) edge (c);
\draw[arc] (b) edge (x) edge (c);
\draw[arc] (c) edge (s3);
\draw[arc] (d) edge (c) edge (f);
\draw[arc] (e) edge (b) edge (c) edge (f) edge (s0) edge (s3);
\draw[arc] (f) edge (s5);
\draw[arc] (g) edge (f) edge (f2);
 
\draw[arc] (x2) edge (s0b);
\draw[arc] (a2) edge (x2) edge (c2);
\draw[arc] (b2) edge (x2) edge (c2);
\draw[arc] (c2) edge (s3b);
\draw[arc] (d2) edge (c2) edge (f2);
\draw[arc] (e2)edge(b2)edge(c2)edge(f2)edge(s0b)edge(s3b);
\draw[arc] (f2) edge (s5b);
 
\draw[arc, dashed] (0.2,6.3) to (x);
\draw[arc, dashed] (0.6,6.1) to (x);
\draw[arc, dashed] (1.0,5.8) to (x);
\draw[arc, dashed] (2.0,5.2) to (s0);
\draw[arc, dashed] (2.5,5.0) to (s0);
 
\draw[arc, dashed] (7.4,6.1) to (x2);
\end{tikzpicture}
 
\caption{\label{fig-hull-var}Gadget of the variable $x_i$. Recall that $S$ is the set of all sinks and sources of $D$, which are represented in black, and that there is an arc from any source of $S$ to any sink of $S$, which are omitted. The dashed arcs represent the only connections with white (not in $S$) vertices outside the gadget.}
\end{figure}
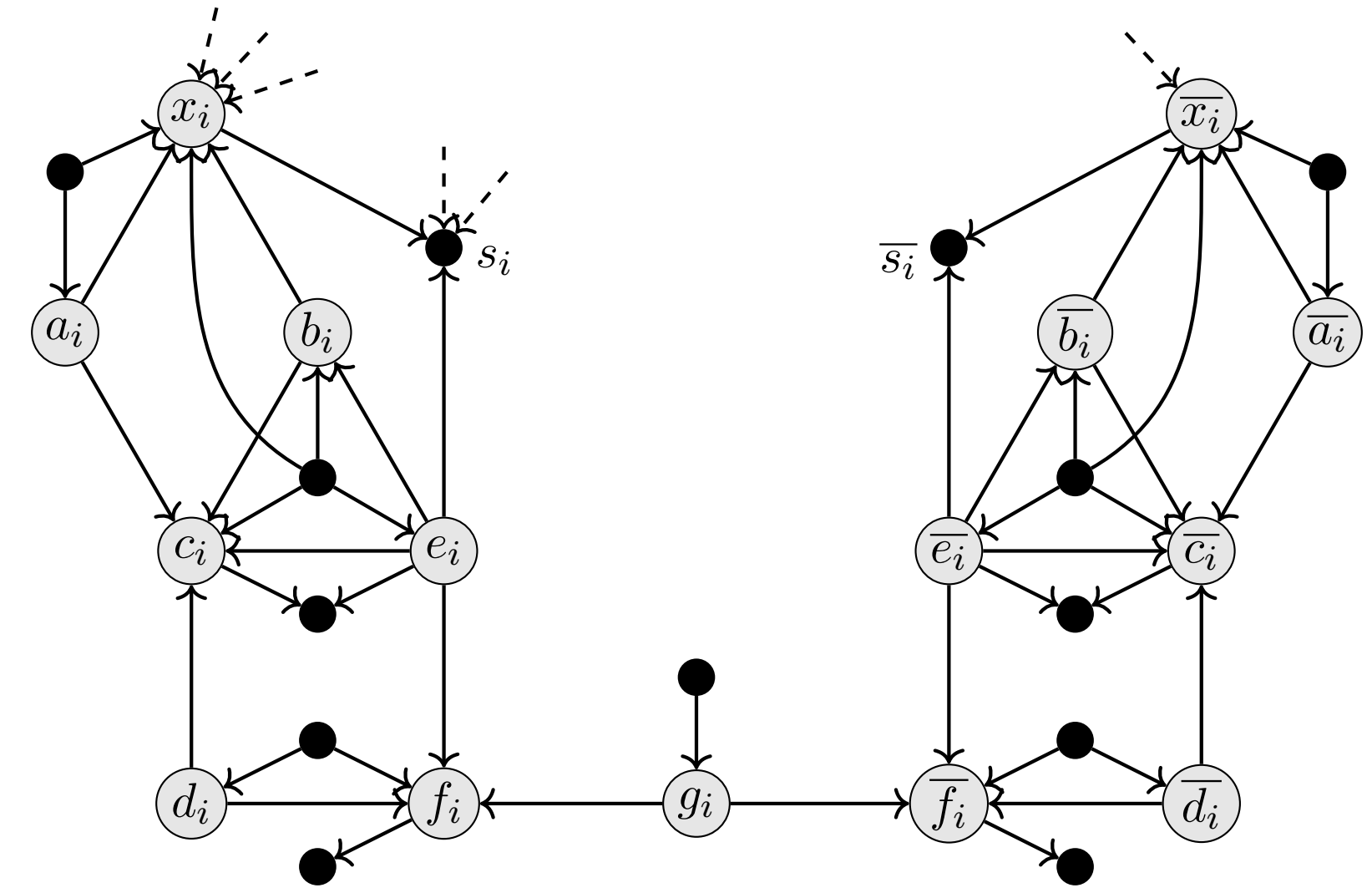

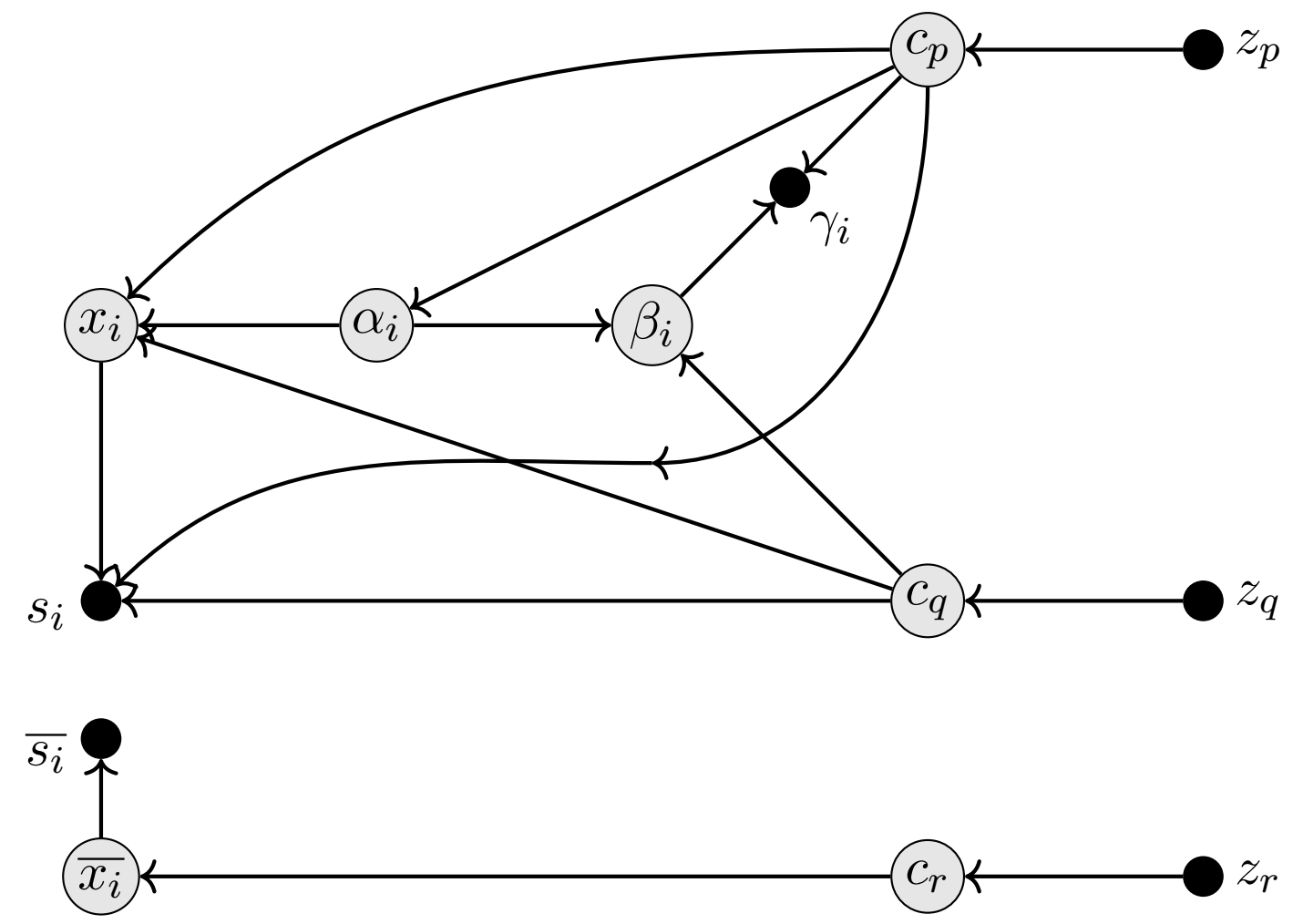
\begin{figure}[ht]\centering
\begin{tikzpicture}[arc/.style={->, thick}]
\tikzstyle{v1}=[draw,circle,fill=black!10,minimum size=15pt,inner sep=1pt]
\tikzstyle{v2}=[draw,circle,fill=black!100,minimum size=8pt,inner sep=0pt]
 
\node[v1] (x) at (0,2) {$x_i$};
\node[v2] (s) at (0,0) {};\node at (-.4,-.1) {$s_i$};
\node[v1] (a) at (2,2) {$\alpha_i$};
\node[v1] (b) at (4,2) {$\beta_i$};
\node[v2] (s1)at (5,3) {};\node at (5.3,2.7) {$\gamma_i$};
\node[v1] (cp)at (6,4) {$c_p$};
\node[v2] (s2)at (8,4) {};\node at (8.4,4) {$z_p$};
\node[v1] (cq)at (6,0) {$c_q$};
\node[v2] (s3)at (8,0) {};\node at (8.4,0) {$z_q$};
 
\draw[arc] (a) edge (x) edge (b) (b) edge (s1);
\draw[arc] (cp) edge (a) edge (s1);
\draw[arc] (s2) edge (cp) (s3) edge (cq) (x) edge (s);
\draw[arc] (cq) edge (x) edge (s) edge (b);
\draw[arc] (cp) to[out=-180,in=45] (x);
\draw[arc] (cp) to[out=-90,in=0] (4,1);
\draw[arc] (4,1) to[out=-180,in=45] (s);
 
\node[v1] (xb) at (0,-2) {$\overline{x_i}$};
\node[v2] (sb) at (0,-1) {};\node at (-.4,-1.1) {$\overline{s_i}$};
\node[v1] (cr) at (6,-2) {$c_r$};
\node[v2] (s4) at (8,-2) {};\node at (8.4,-2) {$z_r$};
\draw[arc] (cr) edge (xb) (xb) edge (sb) (s4) edge (cr);
 
\end{tikzpicture}
 
\caption{\label{fig-hull-clause}Gadgets of the clauses containing the variable $x_i$, assuming that $C_p$ and $C_q$ are the clauses containing $x_i$ and $C_r$ is the clause containing $\overline{x_i}$. The vertices $x_i,\bar{x}_i,s_i$ and $\bar{s}_i$ are those of the gadget of variable $x_i$.}
\end{figure} 
 
One important property of the constructed DAG $D$ is that there is no induced path of size greater than $3$. That is, every induced path is a $P_2$ (arc) or a $P_3$.

This can be seen by analyzing the longest induced directed paths starting from each vertex of Figures \ref{fig-hull-var} and \ref{fig-hull-clause}.
 
We say that a vertex $u\not\in S$ infects a vertex $v$ if there is an induced path $s\to v\to u$ or $u\to v\to s$, where $s\in S$.
We also say that vertices $u,w\not\in S$ infect a vertex $v$ if there is an induced path $u\to v\to w$ or $w\to v\to u$.
 
Notice that the gadget of any variable $x_i$ has $13$ black vertices and the only white vertices with neighbors outside the gadget are $x_i$ and $\overline{x_i}$.
Notice that no vertex inside the variable gadget is infected even if $x_i$, $\overline{x_i}$, all black vertices and all vertices outside the gadget are infected (recall that there is a direct arc from every source to every sink).
Then, in order to infect the whole variable gadget, we need at least one white infected vertex inside the gadget.
 
Suppose that $d_i$ is infected. Then $d_i$ infects $c_i$ and $f_i$; $c_i$ infects $a_i$; $f_i$ infects $e_i$ and $g_i$; $a_i$ infects $x_i$; $g_i$ infects $\overline{f_i}$, which infects $\overline{e_i}$. Finally, $x_i$ and $e_i$ infect $b_i$. No other vertex is infected if $\overline{x_i}$ is not infected. On the other hand, if $\overline{x_i}$ is infected, then $\overline{e_i}$ and $\overline{x_i}$ infect $\overline{b_i}$, which infects $\overline{c_i}$, which infects $\overline{a_i}$ and $\overline{d_i}$.
By symmetry, we have the analogous situation if we had $\overline{d_i}$ infected at the beginning, instead of $d_i$.
 
First suppose that the literal $x_i$ (positive) appears in the clauses $C_p$ and $C_q$. If the vertex $x_i$ is infected, it infects $c_p$ and $c_q$. Then $c_q$ infects $\beta_i$ and later $\beta_i$ and $c_p$ infect $\alpha_i$.
On the other hand, if $c_p$ and $c_q$ are infected, then $c_q$ infects $\beta_i$ and later $\beta_i$ and $c_p$ infect $\alpha_i$, which infects $x_i$.
However, if only one of $c_p$ and $c_q$ is infected and the vertices of the variable gadget of $x_i$ cannot infect $x_i$, then $x_i$ is not infected.
 
Now suppose that the literal $x_i$ or the literal $\overline{x_i}$ (say $\overline{x_i}$) appears only in the clause $C_r$. If the vertex $\overline{x_i}$ is infected, it infects $c_r$. On the other hand, if $c_r$ is infected, it infects $\overline{x_i}$. However, if $c_r$ is not infected and the vertices of the variable gadget of $x_i$ cannot infect $\overline{x_i}$, then $\overline{x_i}$ is not infected.
 
Therefore, let us go to the final arguments.
Assume that there is a truth assignment that satisfies the formula. We will obtain a monophonic hull set $R$ of size $14n+m+\ell$. Add to $R$ all black vertices, which sum $13n+m+\ell$ vertices. If $x_i$ is true, add $d_i$ to $R$; otherwise, add $\overline{d_i}$ to $R$. Thus, $R$ has size $14n+m+\ell$. As shown before, we have that the vertices $x_i$ and $\overline{e_i}$ (resp. $\overline{x_i}$ and $e_i$) are infected if $x_i$ is true (resp. false).
Since it is a truth assignment, then all clause vertices $c_p$ will be infected. Moreover, if $x_i$ is true, then $\overline{x_i}$ is infected by its clause vertex and then $\overline{x_i}$ and $\overline{e_i}$ infect the rest of the variable gadget of $x_i$. Finally, if $x_i$ is false, then $x_i$ is infected by its clause vertices and then $x_i$ and $e_i$ infect the rest of the variable gadget of $x_i$. With this, all vertices are infected and $R$ is a monophonic hull set.
 
Now assume that $D$ has a monophonic hull set $R$ with $14n+m+\ell$ vertices. Then $R$ must contain all black vertices, summing $13n+m+\ell$ vertices (recall that, in every digraph, all sources and sinks belong to any hull set).
As shown before, $R$ must contain at least (in fact, exactly) one white vertex inside the gadget of each variable $x_i$, where we may assume without loss of generality that it is $d_i$ or $\overline{d_i}$.
Moreover, $x_i$ and $\overline{x_i}$ are infected only in two cases: either by vertices inside its gadget or by its clause vertices. Thus, since $R$ is a monophonic hull set, every clause vertex $c_p$ is infected by a vertex $x_i$ or $\overline{x_i}$ which was infected by the vertices inside its gadget.
Thus, by either assigning true if $d_i\in R$ or false if $\overline{d_i}\in R$, all clauses are satisfied.
 
To see that $D$ is a DAG, first note that no vertices $x_i$ or $\bar{x_i}$ are part of a directed cycle since they have a single out-neighbor and this is a sink. Hence, every directed cycle should be fully included either in a variable-gadget or in a clause gadget. It can be checked that there is no such directed cycle.
\end{proof}
 
Since the DAG $D$ of the reduction of the proof above has no induced directed path of size 4 or more, then the following also holds for the $P_3^*$ convexity.
We remark that this argument does not extend to the $P_3$ convexity: since paths on three vertices need not be minimal, the set of sources and sinks of $D$ is already a hull set in the $P_3$ convexity.
 
\begin{corollary}
Given a directed graph $D$ and an integer $k$, deciding if the hull number of $D$ is at most $k$ is $\NP$-complete for the $P_3^*$ convexity, even in DAGs.
\end{corollary}
 
\section{Monophonic hull is polynomial in tournaments}
 
In this section, we study the monophonic directed convexity of tournaments, establishing structural properties and bounds on the hull number and interval number. First, we prove that for any tournament $T$, $\ohnm(T)$ can be computed in linear time in $|V(T)|+|A(T)|$. Second, we show a family of strong tournaments $\mathcal{T} = \{T_1,T_2,T_3,\ldots\}$ such that for each $k \geq 1$, $|V(T_k)| = 4k$ and $\oinm(T_k) = 2k = |V(T_k)|/2$.

A tournament is a directed graph $T = (V,A)$ such that, for every pair of vertices $v,w \in V$, either $(v,w) \in A$ or $(w,v) \in A$, but not both. The underlying graph is complete in a tournament, so the strong monophonic directed convexity is trivial: every vertex subset is convex. This is why only the monophonic convexity is studied in this section.

We denote by ${\cal S}(T)$ the set of strongly connected components of the tournament $T$, i.e., the set of the induced subgraphs $T'$ of $T$ that are maximal with respect to the property that every vertex of $T'$ is reachable from every vertex of $T'$. A tournament $T$ is \emph{strong} if $|{\cal S}(T)|=1$. Every strongly connected component of a tournament is a strong subtournament. A \emph{transitive tournament} is an acyclic tournament. We denote by $TT_n$ the transitive tournament on $n$ vertices. 

First, let us prove a very useful technical lemma.
\begin{lemma}
\label{lem:techtourns}
Let $T$ be a tournament, and $P$ be a minimal path in $T$. If $S$ contains at least two vertices of $P$, then $V(P) \subseteq \ohullm(S)$.
\end{lemma}

\begin{proof}
Let $P = z_1,z_2,\ldots,z_n$ be a minimal path such that $n \geq 2$. Note that, for any $i < j$, every internal vertex of $P$ between $z_i$ and $z_j$ belongs to $\oifm(\{z_i,z_j\})$, since any subpath of a minimal path is also a minimal path.

Furthermore, the predecessor $z_{i-1}$ and successor $z_{i+2}$ of a pair of vertices $z_i$ and $z_{i+1}$ in $P$, assuming they exist, belong to $\oifm(\{z_i,z_{i+1}\})$, since the paths $z_{i+1},z_{i-1},z_i$ and $z_{i+1},z_{i+2},z_i$ are minimal.

So, let $S$ be a set of vertices of $T$ such that  $z_i,z_j \in S$ for some $1 \leq i < j \leq n$. It follows that $z_k \in \oifm(S)$ for every $i < k < j$. After that, if we apply the function $\oifm$ repeatedly, the vertices before $z_i$ and after $z_j$ in $P$ ``enter'' the hull set, i.e., $z_{i-x+1} \in \oi^{\;x}_{\mc}(S)$ for every $2 \leq x \leq i$, and $z_{j+x-1} \in \oi^{\;x}_{\mc}(S)$ for every $2 \leq x \leq n+1-j$. Therefore, we conclude that $V(P) \subseteq \ohullm(S)$.
\end{proof}

\begin{lemma}
\label{lem:sumconcomp}
Let $T$ be a tournament. Then, 
\begin{align*}
\oinm(T)=\;&\sum_{T' \in {\cal S}(T)} \oinm(T'),\text{ and}\\
\ohnm(T)=\;&\sum_{T' \in {\cal S}(T)} \ohnm(T').
\end{align*}
\end{lemma}

\begin{proof}
If $P = u_1,u_2,\ldots,u_k$ is a minimal path of length at least two in a tournament $T$, then there is an arc from $u_k$ to $u_1$. This implies that $C = u_1,u_2,\ldots,u_k,u_1$ is a cycle in $T$. Such a cycle cannot contain a vertex from two distinct strongly connected components in $T$ because, otherwise, they would be a single strongly connected component. Thus, no minimal paths that contain some internal vertex can contain vertices from two distinct strongly connected components.

By this observation, if $u$ and $v$ are vertices of two distinct strongly connected components, $\oifm(\{u,v\}) = \{u,v\}$; and, if $u$ and $v$ are vertices of the same strongly connected component, $\oifm(\{u,v\})$ contains only vertices of the same strongly connected component as them. Thus, for any vertex set $S$ and strongly connected component $Q$, $\oifm(S) \cap V(Q) = \oifm(S \cap V(Q))$, i.e., the vertices in $\oifm(S) \setminus S$ inside a specific strongly connected component $Q$ are in $\oifm(S)$ solely because of the vertices in $S$ that are in the same strongly connected component $Q$.

We conclude that the hull set (resp. interval set) of $T$ is the union of the hull sets (resp. interval sets) of its strongly connected components. The lemma follows.
\end{proof}

Lemma \ref{lem:sumconcomp} is important because it reduces the hull and interval numbers of an arbitrary tournament to those of its strongly connected components, letting us focus on strong tournaments.

\begin{lemma}
\label{lem:strongtourna}
If $T$ is a strong tournament with at least two vertices, any vertex set of size two is a hull set of $T$ in the monophonic directed convexity.
\end{lemma}

\begin{proof}
Let $T = (V,A)$ be a strong tournament with $n \geq 2$ vertices. To prove the lemma, we have to prove that if $S$ is a subset of $V$ of size at least two and at most $n-1$, then $S \subsetneq \oifm(S)$. By Lemma \ref{lem:techtourns}, if there is a minimal path $P$ in $T$ that contains at least two vertices in $S$ and at least one vertex $q$ not in $S$, then $q$ is in $V(P) \subseteq \ohullm(S)$ and is not in $S$. 
This implies that $S \subsetneq \oifm(S)$, since if $S=\oifm(S)$, then $S=\ohullm(S)$, which is a contradiction given that $q \in \ohullm(S)$ and $q \notin S$.

Therefore, our goal is to prove that, if $S \subseteq V$ has size at least two and at most $n-1$, there is a minimal path in $T$ that contains at least two vertices in $S$ and at least one vertex in $S' = V \setminus S$.

First, we claim the following.
\begin{claim}
There is a path $u \to v \to w$ in $T$ such that $u,v \in S$ and $w \in S'$.
\end{claim}

\begin{proofclaim}
Assume that there is a vertex $z \in S$ such that there is no arc from $z$ to any vertex in $S'$. Since $T$ is strong, there is a path between $z$ and any vertex in $S'$. Let $P$ be a path of minimum length over all paths that starts at $z$ and finishes at some vertex of $S'$. Since there is no arc from $z$ to some vertex in $S'$, $P$ has at least three vertices. Thus, the last three vertices $u,v$ and $w$ of $P$ form a path such that $u,v \in S$ and $w \in S'$.

Now, assume that for every $z \in S$, there is an arc from $z$ to some vertex in $S'$. Let $(u,v)$ be any arc of $T[S]$, which exists since $S$ has at least two vertices, and let $w$ be a vertex in $S'$ such that $(v,w)$ is an arc of $T$. Therefore, $u \to v \to w$ is a path in $T$ such that $u,v \in S$ and $w \in S'$.
\end{proofclaim}

\begin{figure}[htbp]
\centering
\begin{tikzpicture}[
    vertex/.style={circle, draw=black, outer sep=1.5pt, minimum size=7mm, inner sep=0pt},
    edge/.style={-{Stealth[length=2.5mm]},black},
    wavy/.style={-{Stealth[length=2.5mm]},black,
                 decorate, decoration={snake, amplitude=0.6mm, segment length=3mm, post length=2mm}},
    wavyu/.style={-,black,
                 decorate, decoration={snake, amplitude=0.6mm, segment length=3mm, post length=1mm}}
    ]
    \node[vertex] (u) at (0,2)   {$u$};
    \node[vertex] (v) at (0,0) {$v$};
    \node[vertex] (w) at (7,0) {$w$};
    \node[vertex] (zi) at (7,2)   {$z_i$};
    \node[vertex] (zip1) at (5.5,2)   {$z_{i+1}$};
    \node (zret) at (3.5,2)   {$\ldots$};
    \node[vertex] (zk) at (1.5,2)   {$z_k$};
    
    \draw[edge] (u) -- (v);
    \draw[edge] (v) -- (w);
    \draw[wavy,color=red!60!black] (w) -- (zi);
    \draw[edge,color=red!60!black] (zi) -- (zip1);
    \draw[wavyu,color=red!60!black] (zip1) -- (zret);
    \draw[wavy,color=red!60!black] (zret) -- (zk);
    \draw[edge,color=red!60!black] (zk) -- (u);
    
    \draw[edge] (zip1) -- (v);
    \draw[edge] (zret) -- (v.45);
    \draw[edge] (zk) -- (v.70);
    \draw[edge] (v.5) -- (zi.240);
\end{tikzpicture}
\caption{\label{fig:pathPintegeri}The figure illustrates the vertices $u,v$ and $w$ and the path $P$ with some of its vertices, including $z_i$, and the arcs to and from $v$. Wavy edges mean a subpath of $P$; and red edges represent the edges of~$P$.}
\end{figure}
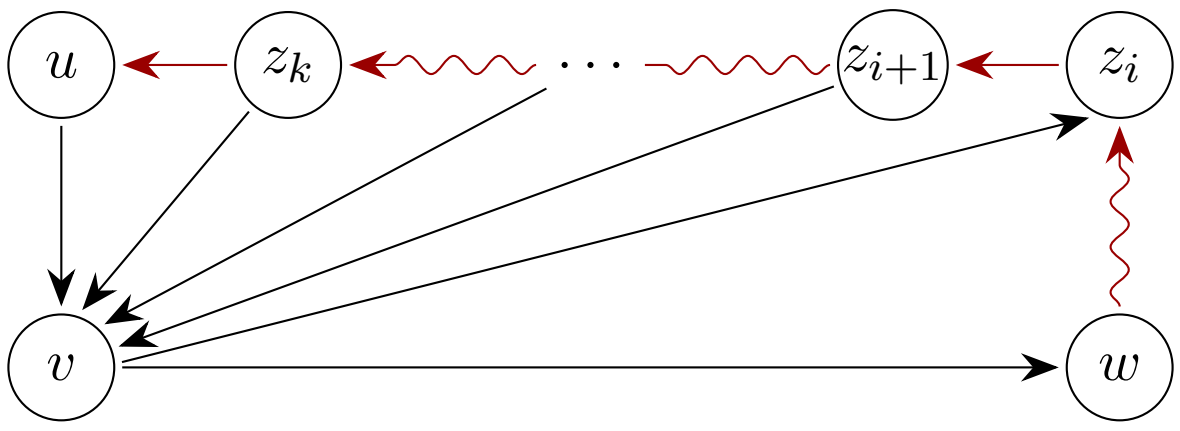

So, let $u \to v \to w$ be a path in $T$ such that $u,v \in S$ and $w \in S'$. Additionally, let $P = (z_0, z_1, \ldots, z_k, z_{k+1})$ be a shortest path, which is therefore minimal, between $w = z_0$ and $u = z_{k+1}$. Note that $k = 0$ when there is an arc from $w$ to $u$. If $v$ is a vertex of $P$, then $P$ is a minimal path containing $u,v \in S$ and $w \in S'$, concluding our proof.

Therefore, assume that $v$ is not a vertex of $P$. Let $i$ be the greatest integer such that there is an arc from $v$ to $z_i$. Since there is an arc from $v$ to $w = z_0$ and from $u = z_{k+1}$ to $v$, it follows that $0 \leq i \leq k$. Figure~\ref{fig:pathPintegeri} illustrates the vertices $u,v$ and $w$ and the path $P$ with some of its vertices, including $z_i$, and the arcs to and from $v$. Wavy edges mean a subpath of $P$; and red edges represent the edges of~$P$.

Since $P$ is minimal, and there is an arc from $v$ to $z_i$ and from $z_j$ to $v$ for every $i+1 \leq j \leq k+1$, we conclude that $P' = (v, z_i, z_{i+1}, \ldots, z_k, u)$ is a minimal path in $T$. If $z_i \in S$, the minimal path $P$ contains $w \in S'$ and $z_i,u \in S$. Otherwise, the minimal path $P'$ contains $z_i \in S'$ and $u,v \in S$.
\end{proof}

\begin{theorem}
For any tournament $T$, $\ohnm(T)$ can be computed in $O(|V(T)| + |A(T)|)$ time.
\end{theorem}

\begin{proof}
Since $\oifm(\{x\}) = \{x\}$ for any vertex $x$ in $T$, by Lemma~\ref{lem:strongtourna}, the hull number of any strongly connected component of $T$ with at least two vertices is exactly $2$. Let $c_1$ be the number of strongly connected components of $T$ with one vertex, and $c_2$ be the number of strongly connected components of $T$ with at least two vertices. By Lemma~\ref{lem:sumconcomp}, we have:
$$\ohnm(T) = c_1 + 2 \cdot c_2$$

Using Kosaraju's Algorithm~\cite{SHARIR1981}, our algorithm computes all strongly connected components of $T$, identifying how many have only one vertex and how many have more than one vertex. Since Kosaraju's Algorithm runs in linear time in the size of its input, the theorem follows.
\end{proof}
 
\begin{theorem}\label{theo:ex}
For each $k \geq 1$, there exists a strong tournament $T_k$ with $n = 4k$ vertices such that $\oinm(T_k) = n/2$.
\end{theorem}
\begin{proof}
Let $Q$ be the tournament with vertex set $\{a,b,c,d\}$ and arc set $\{(a,b)$, $(a,c)$, $(b,c)$, $(b,d)$, $(c,d)$, $(d,a)\}$. In other words, $Q$ is obtained from the transitive tournament $TT_4$ with all arcs oriented according to the order $a<b<c<d$ by reversing the arc between $a$ and $d$. Since $(a,b,c,d,a)$ is a Hamiltonian cycle, $Q$ is strong. No single vertex is an interval set of $Q$, and $\oifm(\{a,d\}) = V(Q)$, as $a,b,d$ and $a,c,d$ are minimal paths in $Q$. Therefore, $\oinm(Q)=2=|V(Q)|/2$.

For any integer $k \geq 1$, let $T_k$ be the tournament with $n=4k$ vertices obtained as follows. Start with $k$ disjoint copies of $Q$. For each $1 \leq i \leq k$, let $Q_i$ be the $i$th copy of $Q$, with vertex set $V_i = \{a_i,b_i,c_i,d_i\}$ and arc set $A_i$ defined in the same way as $Q$. For every $2 \leq i \leq k$, add the arc $(a_i,d_{i-1})$. Now, for every $1 \leq i<j \leq k$ and for every $x \in V_i$ and $y \in V_j$, except for $x=d_i$ and $y=a_{i+1}$, add the arc $(x,y)$. Figure~\ref{fig:exampleinterv} illustrates the tournament $T_3$.

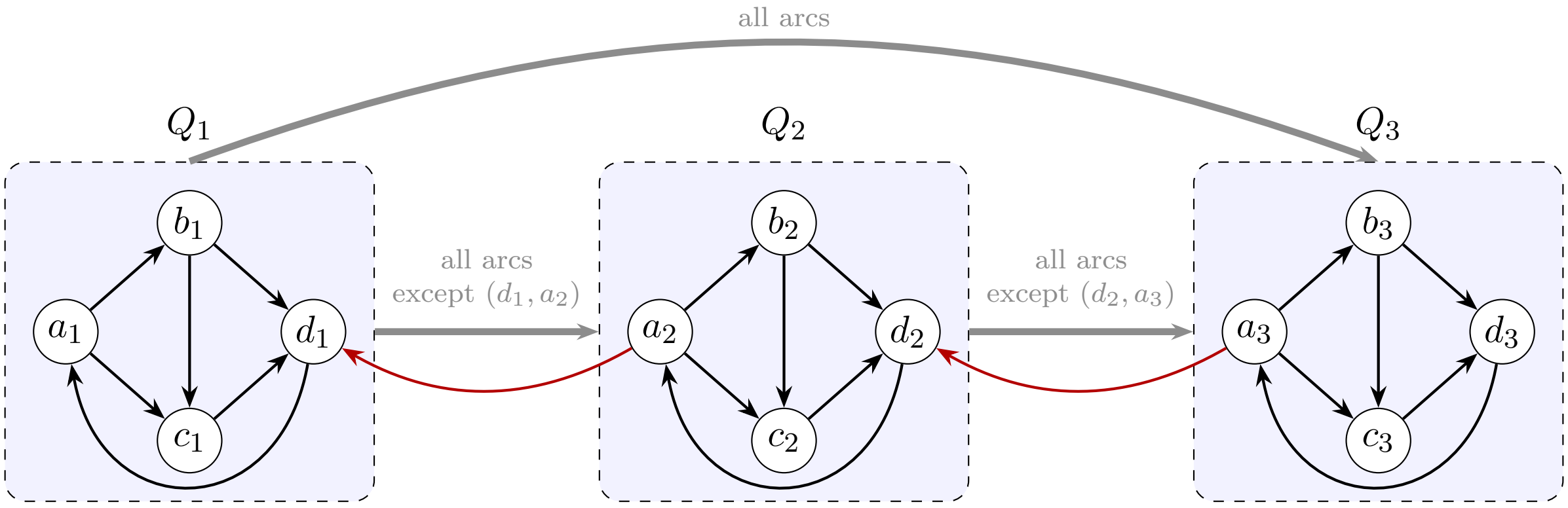
\begin{figure}[htbp]
\centering
\begin{tikzpicture}[
    >={Stealth[length=2.2mm]},
    vtx/.style={circle, draw, inner sep=1.5pt, minimum size=6.5mm, fill=white},
    inarc/.style={->, thick},                               
    backarc/.style={->, thick, red!70!black},               
    fwdarc/.style={->, line width=2pt, gray!90}             
  ]

  \foreach \i/\x in {1/0, 2/6, 3/12} {
    \node[vtx] (a\i) at (\x,0)        {$a_{\i}$};
    \node[vtx] (b\i) at (\x+1.25,1.1)  {$b_{\i}$};
    \node[vtx] (c\i) at (\x+1.25,-1.1) {$c_{\i}$};
    \node[vtx] (d\i) at (\x+2.5,0)    {$d_{\i}$};

    \draw[inarc] (a\i) -- (b\i);
    \draw[inarc] (a\i) -- (c\i);
    \draw[inarc] (b\i) -- (c\i);
    \draw[inarc] (b\i) -- (d\i);
    \draw[inarc] (c\i) -- (d\i);
    \draw[inarc] (d\i) to[bend left=80, looseness=1.8] (a\i);
  }

  \begin{scope}[on background layer]
    \foreach \i in {1,2,3} {
      \node[draw, rounded corners=6pt, fill=blue!5, dashed, inner sep=8pt,
            fit=(a\i)(b\i)(c\i)(d\i)] (box\i) {};
    }
  \end{scope}
  \foreach \i in {1,2,3} {
    \node[above=2pt of box\i.north] {$Q_{\i}$};
  }

  \draw[fwdarc] (box1.east) -- node[above=2pt, font=\scriptsize, align=center]
      {all arcs\\ except $(d_1,a_2)$} (box2.west);
  \draw[fwdarc] (box2.east) -- node[above=2pt, font=\scriptsize, align=center]
      {all arcs\\ except $(d_2,a_3)$} (box3.west);
  \draw[fwdarc] (box1.north) to[bend left=20] node[above, font=\scriptsize]
      {all arcs} (box3.north);

  \draw[backarc] (a2) to[bend left=30] (d1);
  \draw[backarc] (a3) to[bend left=30] (d2);
\end{tikzpicture}
\caption{\label{fig:exampleinterv}The tournament $T_3$, built from three copies $Q_1,Q_2$ and $Q_3$ of the tournament $Q$. A gray arc from $Q_i$ to $Q_j$ means that all the arcs between their vertices are oriented from $Q_i$ to $Q_j$, with the exception of the arcs $(d_1,a_2)$ and $(d_2,a_3)$. These arcs are oriented backwards instead (red arcs $(a_2,d_1)$ and $(a_3,d_2)$).}
\end{figure}

Each $Q_i$ induces a strong tournament, which means that there is a path between any two vertices of $Q_i$. Also, there is an arc from $a_i$ to $a_{i+1}$ and from $a_{i+1}$ to $d_i$, which links consecutive copies of $Q$ in both directions. Therefore, we conclude that $T_k$ is strong for any $k \geq 1$. 

Let $k \geq 1$, and let $T = T_k$ be a tournament with $n = 4k$ vertices. Since $\oifm(\bigcup_{1 \leq i \leq k} \{a_i,d_i\})=V(T)$, $\oinm(T)\leq 2k = n/2$.

Let $S \subseteq V(T)$ be an arbitrary interval set of $T$. To prove that $\oinm(T) \geq 2k = n/2$, we show that $|S \cap V_i|\geq 2$ for every $1 \leq i \leq k$. More precisely, we show that for every $1 \leq i \leq k$, if $\{a_i,d_i\}$ is not a subset of $S$, then $b_i,c_i \in S$.

\begin{claim}
For any $1 \leq i \leq k$, the only minimal paths containing $b_i$ or $c_i$ as internal vertices are $P = a_i,b_i,d_i$ and $P' = a_i,c_i,d_i$.
\end{claim}

\begin{proofclaim}
Let $x \in \{b_i,c_i\}$. By contradiction, let $R \neq P,P'$ be a minimal path containing $x$ as an internal vertex. By the construction of $T_k$ (see Figure~\ref{fig:exampleinterv}), except for $p=a_i$ and $q=d_i$, for any in-neighbor $p$ of $x$ and any out-neighbor $q$ of $x$, $(p,q)$ is an arc, i.e., $(q,p)$ is not an arc. Therefore:
$$\text{($\star$) $p,x,q$ is a minimal path} \implies \text{$p=a_i$ and $q=d_i$.}$$

Since $R$ is minimal, there is an arc from the successor of $x$ in $R$ to the predecessor of $x$ in $R$, which means they form a minimal path with $x$ as an internal vertex. Thus, by $(\star)$, $R = r_1,\ldots, r_s,a_i,x,d_i,r'_1,\ldots,r'_t$, where $s \geq 0$ and $t \geq 0$. However, since $R \neq P,P'$, then $s \geq 1$ or $t \geq 1$. Assume $s \geq 1$. Since $R$ is minimal, $(x,r_s)$ is an arc of $T$, which means that $a_i,x,r_s$ is a minimal path. Therefore, by $(\star)$, $r_s = d_i$, a contradiction since $R$ is a path. The case $t \geq 1$ is analogous.
\end{proofclaim}

Since the only minimal directed paths containing $b_i$ or $c_i$ as internal vertices are $P = a_i,b_i,d_i$ and $P' = a_i,c_i,d_i$, if we assume that $\{a_i,d_i\}$ is not a subset of $S$, we conclude that there are no minimal paths between the vertices of $S$ that contain $b_i$ or $c_i$ as internal vertices. Since $S$ is an interval set of $T$, it follows that $b_i,c_i \in S$. 
\end{proof}

This result\footnote{Note that the construction in Theorem~\ref{theo:ex} can be found in the recent manuscript [Capozzoli, Couto, Junchaya, Interval number for tournaments
in $P_3$-convexity, arXiv, 2026] 
but, as the authors mention on Page~9 and in their reference~[9], they learned about this construction from a talk by one of the authors of the current paper.}
shows that no upper bound on $\oinm(T)$ universally valid for all strong tournaments can be smaller than $|V(T)|/2$ when $|V(T)|$ is divisible by $4$.

The results in this section provide a linear-time algorithm for computing the hull number of a tournament and a family of strong tournaments whose interval number is half their number of vertices. These findings motivate the following open question.

\begin{question}
Is the bound $\oinm(T) \leq \lfloor n/2 \rfloor$ valid for every strong tournament $T$ with $n \geq 4$ vertices?
\end{question}

\section{Monophonic interval is $\NP$-hard in tournaments}
 
In this section, we show that determining whether the monophonic interval number of a tournament is at most two is $\NP$-complete even when restricted to strong tournaments. This contrasts with the hull number, which can be computed in linear time in tournaments. 

Before we present the hardness proof, we need the following lemma.

\begin{lemma}
\label{lem:flexible-sat}
Given a 3-CNF formula $\varphi$, one can construct in linear time a 4-CNF formula $\varphi'$ such that:
\begin{enumerate}
    \item[(i)] $\varphi'$ is satisfiable if and only if $\varphi$ is satisfiable; and
    \item[(ii)] if $\varphi$ is satisfiable, then, for every variable $v$ of $\varphi'$ and every Boolean value $\alpha\in\{\mathsf{T},\mathsf{F}\}$, there is an assignment that satisfies $\varphi'$ and gives the value $\alpha$ to $v$.
\end{enumerate}
In particular, if $\varphi$ is satisfiable, then every literal over a variable of $\varphi'$ is true in some assignment that satisfies $\varphi'$.
\end{lemma}

\begin{proof}
Let $\varphi = C_1 \land \cdots \land C_m$ be a 3-CNF formula over $X = \{x_1,\ldots,x_n\}$. Let $Y = \{y_1,\ldots,y_n\}$ and $x$ be new variables, and let $C'_i$ be the clause obtained from $C_i$ by replacing each $x_j$ with $y_j$. Define
$$\varphi' = \bigwedge_{i=1}^m (C_i \lor x) \land (C'_i \lor \overline{x}),$$
a 4-CNF formula with $2n+1$ variables and $2m$ clauses, clearly constructed in linear time.

If an assignment $B$ satisfies $\varphi'$, then $B$ satisfies every $C_i$ when $B(x) = \mathsf{F}$, and every $C'_i$ when $B(x) = \mathsf{T}$. In both cases, $\varphi$ is satisfiable. Conversely, let $A$ be an assignment that satisfies $\varphi$. Let $B$ be an assignment of $X \cup Y \cup \{x\}$ such that $B(x_j) = A(x_j)$ for every $1 \leq j \leq n$ and $B(x) = \mathsf{F}$, and let $B'$ be one such that $B'(y_j) = A(x_j)$ for every $1 \leq j \leq n$ and $B'(x) = \mathsf{T}$. The values of $B$ on $Y$ and of $B'$ on $X$ are arbitrary. Then $B$ satisfies each $C_i \lor x$ via $C_i$ and each $C'_i \lor \overline{x}$ via $\overline{x}$. Symmetrically, $B'$ satisfies $\varphi'$. This proves (i).

Moreover, since the values of $B$ on $Y$ and of $B'$ on $X$ are arbitrary, and $B(x) \neq B'(x)$, every variable of $\varphi'$ receives each value $\alpha \in \{\mathsf{T},\mathsf{F}\}$ in some assignment that satisfies $\varphi'$, which proves (ii).
\end{proof}

\begin{theorem}\label{theo:IA}
Given a strong tournament $T$, deciding whether $\oinm(T) \leq 2$ is $\NP$-complete.
\end{theorem}

\begin{proof}
The problem is in $\NP$. A certificate for a yes-instance $T$ consists of a set
$S = \{v,u\} \subseteq V(T)$, where $(v,u) \in A(T)$, together with, for each vertex $w \in V(T)\setminus S$, a minimal path from $u$ to $v$ that contains $w$. It can be verified in polynomial time, since minimality of a path can be checked in polynomial time.

To prove the problem $\NP$-hard, we show a polynomial-time reduction from 3\SAT. Let $\varphi$ be an instance of 3\SAT. First, we obtain a 4-CNF formula $\varphi'$ with all the properties of Lemma~\ref{lem:flexible-sat} in linear time.

The idea of the construction is the following. The vertex set of the constructed tournament is partitioned into a sequence of sets called \emph{layers}: one layer for each clause (clause layers), one layer for each variable (value layers), and a few singleton layers, among them $\{s\}$, $\{b\}$ and $\{t\}$. Every minimal path from $s$ to $t$ through $b$ must contain exactly one vertex of each layer, in order, passing through one vertex representing a literal in each clause layer, and one vertex in each value layer, representing a truth value of its variable. Such a path cannot contain both ends of a conflict arc, which goes from a vertex representing a literal to a vertex representing a value that falsifies it. Hence, such a path exists exactly when $\varphi'$ is satisfiable. Lemma~\ref{lem:flexible-sat} ensures that, when this happens, every literal and every value is used by some such path, so that $\{s,t\}$ is an interval set. Moreover, $s$ and $t$ belong to every interval set of size two, so $\oinm(T) \leq 2$ exactly when $\varphi'$ is satisfiable.

Let $\varphi' = C_1\wedge\cdots\wedge C_m$ be a 4-CNF formula over the set of variables $X = \{x_1,\ldots,x_n\}$. Let $N=m+n+6$. We define the tournament $T$ as follows. 

\begin{itemize}
	\item The vertex set of $T$ is partitioned into $N+1$ nonempty sets $V_0,V_1,\ldots,V_N$ called \emph{layers};
	\item $V_0=\{s\}$, $V_1=\{s_1\}$ and $V_2=\{s_2\}$;
	\item for every $1\leq i\leq m$, the layer $V_{i+2}$ contains four vertices, each one representing a literal of $C_i$. These layers are called \emph{clause layers};
	\item $V_{m+3}=\{b\}$;
	\item for every $1\leq i\leq n$, the layer $V_{m+3+i}$ contains two vertices $x_i^{\mathsf{T}}$ and $x_i^{\mathsf{F}}$. These layers are called \emph{value layers};
	\item $V_{N-2}=\{t_2\}$, $V_{N-1}=\{t_1\}$ and $V_N=\{t\}$.
\end{itemize}
For a vertex $v$, we denote by $q_v$ the index $h$ such that $v\in V_h$. We say that the vertex $x_i^\mathsf{F}$ (resp.\ $x_i^\mathsf{T}$) \emph{falsifies} every vertex in the clause layers that represents the literal $x_i$ (resp.\ $\overline{x_i}$). Let $v$ and $w$ be distinct vertices with $q_v \leq q_w$. The arc between them is defined by the following rules:
\begin{itemize}
	\item if $q_w = q_v+1$, then $(v,w)$ is an arc;
	\item if $q_w \geq q_v+2$, then $(w,v)$ is an arc, unless $w$ falsifies $v$, in which case $(v,w)$ is an arc, called a \emph{conflict arc};
	\item if $q_v = q_w$, the arc between $v$ and $w$ is oriented arbitrarily.
\end{itemize}
In other words, every layer dominates the next one, arcs between non-consecutive layers go from the later layer to the earlier one, and the only exceptions are the conflict arcs. Since there is an arc from $t$ to $s$, and, for any vertex $v$ of $T$, there is a path from $s$ to $v$ and from $v$ to $t$, it follows that $T$ is strong. Furthermore, $T$ has $7 + 4m + 2n$ vertices and can be constructed in polynomial time. Figure~\ref{fig:intervaltournconstruction} illustrates the construction.

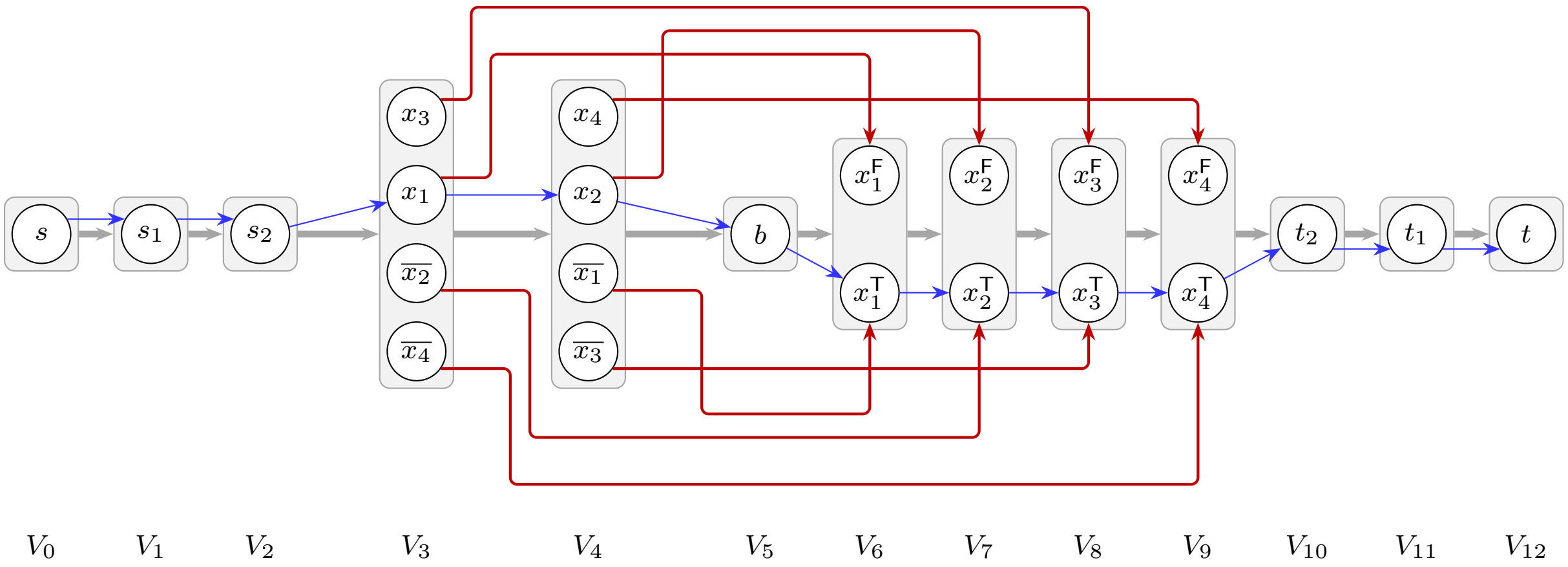
\begin{figure}[htbp]
	\centering
	\begin{tikzpicture}[
		scale=0.8,
		>={Stealth[length=1.8mm]},
		vtx/.style={circle, draw, inner sep=0pt, minimum size=6mm, font=\scriptsize, fill=white},
		set/.style={draw=gray!70, rounded corners=3pt, fill=gray!10, inner sep=2pt},
		arcset/.style={->, gray!70, line width=2pt},
		conf/.style={->, red!75!black, thick, rounded corners=2pt},
		sol/.style={->, blue!80!white}
		]

		\node[vtx] (s)  at (0,0)   {$s$};
		\node[vtx] (s1) at (1.4,0) {$s_1$};
		\node[vtx] (s2)  at (2.8,0) {$s_2$};        
		\node[vtx] (A3) at (4.8,1.5)  {$x_3$};
        \node[vtx] (A1) at (4.8,0.5)  {$x_1$};
        \node[vtx] (A2) at (4.8,-0.5) {$\overline{x_2}$};
        \node[vtx] (A4) at (4.8,-1.5) {$\overline{x_4}$};
        \node[vtx] (B3) at (7.0,1.5)  {$x_4$};
        \node[vtx] (B1) at (7.0,0.5)  {$x_2$};
        \node[vtx] (B2) at (7.0,-0.5) {$\overline{x_1}$};
        \node[vtx] (B4) at (7.0,-1.5) {$\overline{x_3}$};
        \node[vtx] (b)  at (9.2,0) {$b$};
        \foreach \i/\x in {1/10.6,2/12.0,3/13.4,4/14.8}{
            \node[vtx] (F\i) at (\x,0.75)  {$x_{\i}^{\mathsf{F}}$};
            \node[vtx] (T\i) at (\x,-0.75) {$x_{\i}^{\mathsf{T}}$};
        }
		\node[vtx] (t2)  at (16.2,0) {$t_2$};
		\node[vtx] (t1) at (17.6,0) {$t_1$};
		\node[vtx] (t)  at (19.0,0) {$t$};
		
        \begin{scope}[on background layer]
        \foreach \nm/\members in {Xs/(s), Xs1/(s1), Xs2/(s2), XL1/(A3)(A4), XL2/(B3)(B4), Xb/(b), XR1/(T1)(F1), XR2/(T2)(F2), XR3/(T3)(F3), XR4/(T4)(F4), Xt2/(t2), Xt1/(t1), Xt/(t)}
          \node[set, fit=\members] (\nm) {};
        \end{scope}
        
        \foreach \x/\h in {0/0,1.4/1,2.8/2,4.8/3,7.0/4,9.2/5,10.6/6,12.0/7,13.4/8,14.8/9,16.2/10,17.6/11,19.0/12}
        \node[font=\scriptsize] at (\x,-4.0) {$V_{\h}$};
        
        \foreach \u/\w in {Xs/Xs1, Xs1/Xs2, Xs2/XL1, XL1/XL2, XL2/Xb, Xb/XR1, XR1/XR2, XR2/XR3, XR3/XR4, XR4/Xt2, Xt2/Xt1, Xt1/Xt}
            \draw[arcset] (\u.east) -- (\w.west);
		
		\draw[conf] (A1.35) -| (5.75,2.3)  -| (F1.north);  
        \draw[conf] (B1.35) -| (7.95,2.6)  -| (F2.north);  
        \draw[conf] (A3.35) -| (5.5,2.9)   -| (F3.north);  
        \draw[conf] (B3.35) -| (F4.north);  
        \draw[conf] (B2.-35) -| (8.45,-2.3) -| (T1.south);  
        \draw[conf] (A2.-35) -| (6.25,-2.6) -| (T2.south);  
        \draw[conf] (B4.-35) -| (T3.south);  
        \draw[conf] (A4.-35) -| (6.0,-3.2)  -| (T4.south);  
		
		\draw[sol] (s.30) -- (s1.150); \draw[sol] (s1.30) -- (s2.150); \draw[sol] (s2) -- (A1);
		\draw[sol] (A1) -- (B1); \draw[sol] (B1) -- (b); \draw[sol] (b) -- (T1);
		\draw[sol] (T1) -- (T2); \draw[sol] (T2) -- (T3); \draw[sol] (T3) -- (T4);
		\draw[sol] (T4) -- (t2); \draw[sol] (t2.-30) -- (t1.-150); \draw[sol] (t1.-30) -- (t.-150);
	\end{tikzpicture}
	\caption{\label{fig:intervaltournconstruction}The tournament $T$ for the 4-CNF formula $\varphi'=(x_1 \lor \overline{x_2} \lor x_3\lor \overline{x_4}) \land (\overline{x_1} \lor x_2 \lor \overline{x_3} \lor x_4)$. Each gray box is one of the layers $V_i$, and a gray arrow from one layer to the next means that every vertex of the first layer dominates every vertex of the second. Arcs inside a layer and arcs between non-consecutive layers are not drawn. The latter all go from the later layer to the earlier one, except the conflict arcs (red), which go from a vertex in a clause layer to a vertex in a value layer that falsifies it. The blue path is a minimal path from $s$ to $t$ through $b$ associated with the satisfying assignment in which every variable is true.}
\end{figure}

In a tournament, a path $(z_1,z_2,\ldots,z_k)$ is minimal if and only if $(z_q,z_{q'})$ is an arc for every $1 \leq q' \leq q-2$, i.e., every vertex of the path dominates all the vertices that appear at least two positions before it. We use this fact without mentioning it further. 

We say that a path $(z_0,z_1,\ldots,z_k)$ is \emph{monotone} if $k = N$ and $z_q \in V_q$ for every $0 \leq q \leq k$. Note that, in particular, any monotone path goes through $s,s_1,s_2,b,t_2,t_1$ and $t$. Figure~\ref{fig:intervaltournconstruction} illustrates a monotone path in blue, which is also a minimal path.

Suppose $\varphi'$ is satisfiable. Let us prove that $S = \{s,t\}$ is a monophonic interval set of $T$, which implies that $\oinm(T) \leq 2$. By Lemma~\ref{lem:flexible-sat}(ii), every variable receives each Boolean value in some assignment that satisfies $\varphi'$. Hence, the following claim implies that $\oifm(S)=V(T)$.

\begin{claim}
Let $A$ be an assignment that satisfies $\varphi'$, and let $w$ be a vertex in a clause layer that represents a literal that evaluates to true according to $A$. Then there is a minimal path from $s$ to $t$ that contains $w$ and the vertex $x_i^{A(x_i)}$ for every $1 \leq i \leq n$.
\end{claim}

\begin{proofclaim}
Let $P = (z_0,z_1,\ldots,z_k)$ be a monotone path such that:
\begin{itemize}
    \item $z_0=s,z_1=s_1,z_2=s_2,z_{3+m}=b,z_{k-2}=t_2,z_{k-1}=t_1$ and $z_k=z_N=t$;
    \item for each clause layer $V_q$, if $w \in V_q$, let $z_q = w$; otherwise, let $z_q$ be a vertex in $V_q$ that represents a literal that evaluates to true according to $A$, which exists because $A$ satisfies $\varphi'$.
    \item for each value layer $V_q$, let $z_q = x_{q - (m+3)}^d$, where $d = A(x_{q - (m+3)})$.
\end{itemize}
Since $A$ satisfies $\varphi'$, $P$ is well-defined, and, by its definition, it goes through $w$ and $x_i^{A(x_i)}$ for every $1 \leq i \leq n$. By the definition of $P$, there is no conflict arc between two vertices of $P$. Thus, since $P$ is monotone, any two non-consecutive vertices of $P$ are in non-consecutive layers, so the arc between them goes from the later one to the earlier one. Hence, $P$ is minimal.
\end{proofclaim}

We now prove the converse. Assume that $\varphi'$ is not satisfiable. First, let us prove the following claim.

\begin{claim}
If $\oifm(\{s,t\}) \neq V(T)$, then $\oinm(T) > 2$.
\end{claim}

\begin{proofclaim}
Suppose that $\oinm(T) \leq 2$. Clearly, no set of at most one vertex is a monophonic interval set of $T$. Let $S$ be a monophonic interval set of $T$ of size $2$.

For every $2 \leq i \leq N$ and $w \in V_i$, there is an arc from $w$ to $s$, and the only out-neighbor of $s$ is $s_1$. Thus, $(s_2,s,s_1)$ is the only minimal path in $T$ that contains $s$ as an internal vertex. Hence, if $s \notin S$, then $s_1,s_2 \in S$. By similar reasoning, if $t \notin S$, then $t_1,t_2 \in S$. Since $|S| = 2$ and $\{s,s_1,s_2\} \cap \{t,t_1,t_2\} = \emptyset$, this implies that $S = \{s,t\}$.
\end{proofclaim}

Let $S = \{s,t\}$. By this claim, it is enough to prove that $\oifm(S) \neq V(T)$. To do that, let us prove that $b \notin \oifm(S)$. If $P$ is a monotone path, let $A_P$ be the assignment $A_P : X \to \{\mathsf{T},\mathsf{F}\}$ of values for the variables of $\varphi'$ such that $A_P(x_i) = d \in \{\mathsf{T},\mathsf{F}\}$ if $P$ goes through the vertex $x_i^d$ in the value layer $V_{m+3+i}$. 

Let $P = (z_0=s,z_1,\ldots,t=z_k)$ be a minimal path from $s$ to $t$, and assume, by contradiction, that $b \in V(P)$. First, let us prove that $P$ cannot be a monotone path.

\begin{claim}
If $P$ is a monotone path that is also a minimal path, then $A_P$ satisfies $\varphi'$.
\end{claim}

\begin{proofclaim}
Since $P$ is monotone, for each $1 \leq i \leq m$, $P$ contains a vertex of $V_{i+2}$, which represents some literal $\ell_i$ of $C_i$. Since $P$ is a minimal path, no conflict arc joins two vertices of $P$. This means that $A_P$ evaluates each literal $\ell_i$ to true, which implies that $A_P$ satisfies $\varphi'$.
\end{proofclaim}

Therefore, since $\varphi'$ is not satisfiable, $P$ cannot be a monotone path. Let $q$ be the greatest integer such that, for every $0 \leq i \leq q$, $z_i \in V_i$, and let $q'$ be the least integer such that, for every $0 \leq i \leq k-q'$, $z_{k-i} \in V_{N-i}$. Since $s_1$ is the only out-neighbor of $s$, it follows that $q \geq 1$.

Suppose $(z_q,z_{q+1})$ is not a conflict arc. Then, $z_{q+1} \in V_j$ for some $j \leq q-2$ or $j=q$. This implies that there is an arc from $z_{j-1}$ to $z_{q+1}$, a contradiction, since $P$ is minimal. By analogous reasoning, $(z_{q'-1},z_{q'})$ is also a conflict arc. 

Since a conflict arc goes from a vertex in a clause layer to a vertex in a value layer, $z_q$ and $z_{q'-1}$ belong to clause layers, and $z_{q+1}$ and $z_{q'}$ belong to value layers. Hence, $z_0,\ldots,z_q$ are in $V_0,\ldots,V_q$ respectively with $q \leq m+2$, and $z_{q'},\ldots,z_k$ are in layers of index at least $m+4$. Since $b \in V(P)$, it follows that $q+1 < q'$, and $b$ is an internal vertex of the subpath $P'$ of $P$ from $z_{q+1}$ to $z_{q'-1}$, which is minimal since $P$ is minimal.

If $(z_{q'-1},z_{q+1})$ is an arc, then it is a conflict arc. Hence, since every vertex in a clause layer has only one outgoing conflict arc, $z_{q+1} = z_{q'}$. This is a contradiction, since $q+1 \neq q'$ and $P$ is a path. On the other hand, if $(z_{q+1}, z_{q'-1})$ is an arc, then $P'$ is not minimal, again a contradiction.

Thus, no minimal path from $s$ to $t$ contains $b$. Since $(t,s)$ is an arc, $(t,s)$ is the only minimal path from $t$ to $s$, and it does not contain $b$. Therefore, $b \notin \oifm(S)$.
\end{proof}
 
\section{Monophonic interval is $\NP$-hard in bipartite oriented graphs}
 
In this section, we prove that the interval number of the monophonic convexity is $\NP$-hard.
The complexity of the interval number of the strong monophonic convexity is still open. 
 
\begin{theorem}\label{thm:inm-bip}
Deciding if $\oinm(D)\leq 2$ is $\NP$-complete in bipartite oriented graphs $D$.
\end{theorem}
 
\begin{proof}
The problem is in $\NP$: a set $S$ with $|S|\leq 2$ together with, for every $v\notin S$, a minimal directed path between two vertices of $S$ containing $v$ is a polynomial certificate, since it can be checked in polynomial time whether a given directed path is minimal.
We reduce from the 2-\textsc{linkage} problem, which has as input a digraph $D'$ and four distinct vertices $s_1,t_1,s_2,t_2$ of $D'$, and asks whether there exist vertex-disjoint directed paths from $s_1$ to $t_1$ and from $s_2$ to $t_2$. This problem is $\NP$-complete~\cite{FHW1980}. We may assume that $D'$ has at least one arc, since otherwise the answer is trivially \textsc{no}.
 
Let $D'_1$ be obtained from $D'$ by adding a new vertex $c$ and the arcs $(t_1,c)$ and $(c,s_2)$. Let $D'_2$ be obtained from $D'_1$ by subdividing every arc exactly once, that is, every arc $(u,v)$ of $D'_1$ is replaced by a new vertex $w_{uv}$ and the arcs $(u,w_{uv})$ and $(w_{uv},v)$. Let $A=V(D'_1)$ and let $B$ be the set of subdivision vertices. Let $p=w_{t_1c}$ and $q=w_{cs_2}$, and let $A'=A\setminus\{c\}$ and $B'=B\setminus\{p,q\}$. Note that $B'\neq\emptyset$, since $D'$ has an arc. Clearly, $D'_2$ is an oriented graph whose underlying graph is bipartite with parts $A$ and $B$.
Moreover, every directed path of $D'_2$ is minimal, since every subdivision vertex has a unique in-neighbor and a unique out-neighbor.

\begin{figure}[ht]\centering
\begin{tikzpicture}[arc/.style={->, thick}, scale=0.9, every node/.style={transform shape}]
\tikzstyle{v1}=[draw,circle,fill=black!10,minimum size=15pt,inner sep=1pt]
\tikzstyle{v3}=[draw,circle,fill=white,minimum size=13pt,inner sep=0.5pt]
\tikzstyle{v2}=[draw,circle,fill=black!100,minimum size=12pt,inner sep=0pt]
 
\draw[rounded corners,fill=black!3] (3.5,-1.25) rectangle (12.5,1.25);
\node at (8.2,0.85) {\LARGE $D'_2$};
 
\node[v1] (s1) at (4.4,0) {$s_1$};
\node[v1] (t1) at (6.4,0) {$t_1$};
\node[v3] (p)  at (7.3,0) {$p$};
\node[v1] (c)  at (8.2,0) {$c$};
\node[v3] (q)  at (9.1,0) {$q$};
\node[v1] (s2) at (10.0,0) {$s_2$};
\node[v1] (t2) at (12.0,0) {$t_2$};
 
\draw[arc,dashed] (s1) to[bend left=35] node[above] {\small $P_1$} (t1);
\draw[arc,dashed] (s2) to[bend left=35] node[above] {\small $P_2$} (t2);
\draw[arc] (t1)--(p); \draw[arc] (p)--(c); \draw[arc] (c)--(q); \draw[arc] (q)--(s2);
 
\node[v2] (x) at (0,0) {}; \node at (0,0.7) {\LARGE $x$};
\node[v1] (r)  at (1.8,0) {$x'_B$};
\node[v1] (xA) at (1.8,2.3) {$x_B$};
\node[v1] (x1) at (1.0,-2.3) {$x''_B$};
\node[v1] (xB) at (2.6,-2.3) {$x_A$};
 
\draw[arc] (x)--(r); \draw[arc] (r)--(s1);
\draw[arc] (x)--(xA); \draw[arc] (x)--(x1); \draw[arc] (x1)--(xB);
 
\node[v2] (y) at (16,0) {}; \node at (16,0.7) {\LARGE $y$};
\node[v1] (rp) at (14.2,0) {$y'_B$};
\node[v1] (yA) at (14.2,2.3) {$y_B$};
\node[v1] (yB) at (13.4,-2.3) {$y_A$};
\node[v1] (y1) at (15.0,-2.3) {$y''_B$};
 
\draw[arc] (t2)--(rp); \draw[arc] (rp)--(y);
\draw[arc] (yA)--(y); \draw[arc] (yB)--(y1); \draw[arc] (y1)--(y);
 
\draw[arc,line width=1.6pt] (xA) to[out=0,in=90] node[pos=0.55,above] {$A'$} (6.0,1.25);
\draw[arc,line width=1.6pt] (10.4,1.25) to[out=90,in=180] node[pos=0.45,above] {$A'$} (yA);
\draw[arc,line width=1.6pt] (xB) to[out=0,in=-90] node[pos=0.55,below] {$B'$} (6.0,-1.25);
\draw[arc,line width=1.6pt] (10.4,-1.25) to[out=-90,in=180] node[pos=0.45,below] {$B'$} (yB);
 
\node at (8.2,2.4) {\huge $D$};
\end{tikzpicture}
\caption{\label{fig-inm-bip}Oriented graph $D$ with unique source and sink $x$ and $y$. $D'_2$ is obtained from $D'$ by adding $c$ and the arcs $(t_1,c)$, $(c,s_2)$, and by subdividing every arc. Arc label $A'/B'$ is related to any vertex of $A'=A\setminus\{c\}$ or $B'=B\setminus\{p,q\}$. $P_1$ and $P_2$ represent a possible solution of 2-\textsc{linkage}.}
\end{figure}
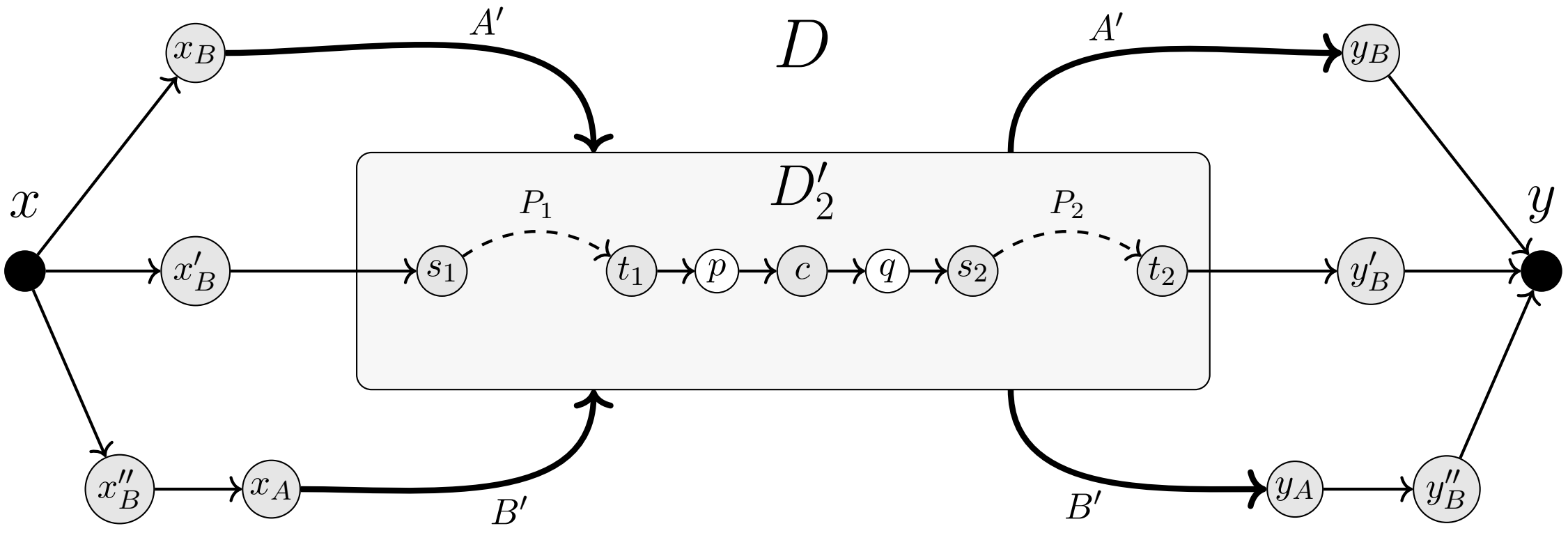
 
Notice that there is a directed path from $s_1$ to $t_2$ in $D'_2$ containing $c$ if and only if $(D',s_1,t_1,s_2,t_2)$ is a \textsc{yes}-instance of 2-\textsc{linkage}.
Indeed, if $P_1$ and $P_2$ are vertex-disjoint directed paths of $D'$ from $s_1$ to $t_1$ and from $s_2$ to $t_2$, then $P_1\,c\,P_2$ is a directed path of $D'_1$, and its subdivision is a directed path of $D'_2$ from $s_1$ to $t_2$ containing $c$. Conversely, a directed path of $D'_2$ from $s_1$ to $t_2$ containing $c$ corresponds to a directed path $P$ of $D'_1$ from $s_1$ to $t_2$ containing $c$. Since $t_1$ is the unique in-neighbor and $s_2$ is the unique out-neighbor of $c$ in $D'_1$, the subpaths of $P$ from $s_1$ to $t_1$ and from $s_2$ to $t_2$ are vertex-disjoint directed paths of $D'$.
 
Now let $D$ be obtained from $D'_2$ by adding ten new vertices $x,y,x_A,x_B,x'_B,x''_B,y_A,y_B,y'_B,y''_B$  (see Figure~\ref{fig-inm-bip}) and the arcs
\begin{itemize}
\item $(x,x_B)$,  $(x,x'_B)$, $(x,x''_B)$, $(x''_B,x_A)$, $(x'_B,s_1)$, $(t_2,y'_B)$, $(y_A,y''_B)$, $(y_B,y)$, $(y'_B,y)$, $(y''_B,y)$;
\item $(x_B,v)$ and $(v,y_B)$ for every $v\in A'$;
\item $(x_A,w)$ and $(w,y_A)$ for every $w\in B'$.
\end{itemize}
Clearly, $D$ is an oriented graph, with bipartition $A\cup\{x,y,x_A,y_A\}$ and $B\cup\{x_B,y_B,x''_B,y''_B,x'_B,y'_B\}$. Moreover, $x$ is the unique source and $y$ is the unique sink of $D$. Since sources and sinks belong to every monophonic interval set, note that $\oinm(D)\leq 2$ if and only if $\oifm(\{x,y\})=V(D)$.
The vertices $x_A/y_A$ (resp.\ $x_B/y_B$) play the role of ``chord makers'': since they are adjacent to every vertex of $B'$ (resp.\ $A'$), a minimal path containing one of them contains at most one vertex of $B'$ (resp.\ $A'$).
 
Notice that $V(D)\setminus\{c,p,q\}\subseteq\oifm(\{x,y\})$, since the following directed paths are minimal: $(x,x_B,v,y_B,y)$ for every $v\in A'$,
$(x,x''_B,x_A,w,y_A,y''_B,y)$ for every $w\in B'$,
$(x,x'_B,s_1,y_B,y)$ and $(x,x_B,t_2,y'_B,y)$.
These paths cover all vertices of $D$, except $c$, $p$ and $q$.
 
Since $t_1$ (resp.\ $c$) is the unique in-neighbor of $p$ (resp.\ $q$) and $c$ (resp.\ $s_2$) is its unique out-neighbor, every path containing (internally) one of $p,c,q$ contains the subpath $(t_1,p,c,q,s_2)$. Hence, $\oifm(\{x,y\})=V(D)$ if and only if there exists a minimal directed path from $x$ to $y$ in $D$ containing $c$.
In the following, we prove that there is a minimal directed path from $x$ to $y$ in $D$ containing $c$ if and only if there is a directed path from $s_1$ to $t_2$ in $D'_2$ containing $c$.
 
Let $Q$ be a directed path from $s_1$ to $t_2$ in $D'_2$ containing $c$ and let $P=(x,x'_B)\,Q\,(y'_B,y)$. As noticed above, $Q$ has no chord. Also, the only arcs between $\{x,x'_B,y'_B,y\}$ and $V(Q)$ are $(x'_B,s_1)$ and $(t_2,y'_B)$, and the only arcs inside $\{x,x'_B,y'_B,y\}$ are $(x,x'_B)$ and $(y'_B,y)$. So $P$ is minimal.
 
Conversely, let $P$ be a minimal directed path from $x$ to $y$ in $D$ containing $c$. Then $P$ contains the subpath $(t_1,p,c,q,s_2)$, with $t_1,s_2\in A'$. Moreover, the successor of $s_2$ in $P$ is either $y_B$ or a vertex $w_{s_2v}\in B'$ (note that $w_{s_2v}\notin\{p,q\}$).
 
First, suppose that the predecessor of $y$ in $P$ is $y_B$. Then the predecessor of $y_B$ is a vertex of $A'$, and one of $t_1,s_2$ is a vertex of $A'$ appearing earlier in $P$, which is adjacent to $y_B$. This is a chord, a contradiction.
 
Now suppose that $P$ ends with $(w,y_A,y''_B,y)$, with $w\in B'$. Since the successor of $s_2$ is a vertex of $B'$ (it is not $y_B$, otherwise $P$ would end with $(y_B,y)$), and every vertex of $B'$ is adjacent to $y_A$, this successor must be $w$. Let us look at the beginning of $P$. If $P$ starts with $(x,x_B)$, then $s_2$ is not the successor of $x_B$ and $(x_B,s_2)$ is a chord. If $P$ starts with $(x,x''_B,x_A,w')$, with $w'\in B'$, then $w'\neq w$ and $(x_A,w)$ is a chord. If $P$ starts with $(x,x'_B,s_1)$, then, since $s_1\neq t_1$, the subpath of $P$ from $s_1$ to $t_1$ contains a vertex of $B'$ (the out-neighbor of $s_1$ in $P$), which is adjacent to $y_A$, a chord. In all cases we get a contradiction.
 
Therefore, $P$ ends with $(t_2,y'_B,y)$. Again, $P$ cannot start with $(x,x_B)$, since then $(x_B,t_2)$ would be a chord, and it cannot start with $(x,x''_B,x_A)$, since then the out-neighbor of $s_2$ in $P$ would be a vertex of $B'$ different from the successor of $x_A$, a chord. Thus $P$ starts with $(x,x'_B,s_1)$. Finally, the subpath of $P$ from $s_1$ to $t_2$ contains none of the new vertices: no vertex of $D'_2$ has an arc to $x,x_B,x_A,x''_B,x'_B$, and entering $y_B$, $y_A$ or $y'_B$ before $t_2$ would force $P$ to reach $y$ before $t_2$. Hence this subpath is a directed path of $D'_2$ from $s_1$ to $t_2$ containing $c$.
 
Therefore, $\oinm(D)\leq 2$ if and only if $(D',s_1,t_1,s_2,t_2)$ is a \textsc{yes}-instance of 2-\textsc{linkage}. Since $D$ can be constructed in polynomial time, the result follows.
\end{proof}

\section{Further research}

In this paper, we started the investigation of the monophonic and the strong monophonic directed convexities. Several questions remain open. Regarding the strong monophonic convexity, the complexity of the interval number is still open. Since the hull number is $\NP$-complete in DAGs for both convexities, it would be interesting to find classes of directed graphs in which they can be computed in polynomial time, as we did for tournaments in the monophonic convexity.

Regarding the monophonic convexity, besides the question about strong tournaments of Section~4, it is natural to consider classes related to tournaments, such as semicomplete digraphs and multipartite tournaments, which were investigated for the $P_3$ convexity~\cite{PWW2006,PARKER2008641,PWW2009}. In particular, we proved that the interval number is $\NP$-complete in bipartite oriented graphs, but the complexity of the hull number in bipartite oriented graphs (and in bipartite tournaments) is open. Finally, other convexity parameters, such as the Carath\'eodory, Helly and Radon numbers and the rank, as well as the parameterized complexity of the hull and interval numbers, have not been investigated for directed monophonic convexities.

\section{Acknowledgements}
Most of the results were obtained by the end of 2025, without the use of AI. An exception is Theorem~\ref{theo:IA}, which was obtained with the assistance of Claude; its main contribution was to suggest the addition of vertex $b$ in the reduction. We also used Claude to produce the TikZ code of the figures.

\bibliographystyle{acm}
\bibliography{mono-dir-conv}

@article{SHARIR1981,
    author = {M. Sharir},
    title = {A strong-connectivity algorithm and its applications in data flow analysis},
    journal = {Computers \& Mathematics with Applications},
    volume = {7},
    number = {1},
    pages = {67--72},
    year = {1981}
}

@article{BHT2012,
  author  = {J{\o}rgen Bang-Jensen and Fr{\'e}d{\'e}ric Havet and Nicolas Trotignon},
  title   = {Finding an induced subdivision of a digraph},
  journal = {Theoretical Computer Science},
  volume  = {443},
  pages   = {10--24},
  year    = {2012}
}

@inproceedings{FLMGTP2026,
  author    = {Florent Foucaud and Lucas Lorieau and Morteza Mohammad-Noori and
               Narges Ghareghani and Prafullkumar Tale and {Parvini Oskuei}, Rasa},
  title     = {Algorithms and Hardness for Geodetic Set on Tree-like Digraphs},
  booktitle = {Algorithms and Discrete Applied Mathematics (CALDAM 2026)},
  series    = {Lecture Notes in Computer Science},
  publisher = {Springer},
  year      = {2026},
  doi       = {10.1007/978-3-032-17156-6_14},
  note      = {arXiv:2603.23193}
}

@article{BGG2009,
  author  = {Paul Balister and Stefanie Gerke and Gregory Gutin},
  title   = {Convex sets in acyclic digraphs},
  journal = {Order},
  volume  = {26},
  pages   = {95--100},
  year    = {2009},
  doi     = {10.1007/s11083-009-9109-9}
}

@article{BGGJRSSY2009,
  author  = {Paul Balister and Stefanie Gerke and Gregory Gutin and Adrian Johnstone
             and Joseph Reddington and Elizabeth Scott and Arezou Soleimanfallah
             and Anders Yeo},
  title   = {Algorithms for generating convex sets in acyclic digraphs},
  journal = {Journal of Discrete Algorithms},
  volume  = {7},
  number  = {4},
  pages   = {509--518},
  year    = {2009}
}

@misc{PAGDT2026,
      title={Convex geometries and directed paths on three vertices}, 
      author={Júlio Araújo and Mitre C. Dourado and Marisa Gutierrez and Fabio Protti and Silvia Tondato},
      year={2026},
      eprint={2606.24707},
      archivePrefix={arXiv},
      primaryClass={math.CO},
      url={https://arxiv.org/abs/2606.24707}, 
}

@article{SCSgrids2017,
  author    = {Gabriela Araujo-Pardo and C{\'e}sar Hern{\'a}ndez-Cruz and Juan Jos{\'e} Montellano-Ballesteros},
  title   = {The Strong Convexity Spectra of Grids},
  journal = {Graphs and Combinatorics},
  volume    = {33},
  number    = {4},
  pages     = {689--713},
  year      = {2017},
}

@book{araujo-livro25,
title = {{I}ntroduction to {G}raph {C}onvexity: an algorithmic approach},
author = {Araújo, Júlio and Dourado, Mitre C and Protti, Fábio and Sampaio, Rudini M.},
year = {2025},
publisher = {Springer Cham},
}

@incollection{jamison1982,
  title={A perspective on abstract convexity: classifying alignments by varieties},
  author={Jamison-Waldner, Robert E.},
  booktitle={Convexity and Related Combinatorial Geometry},
  series={Lecture Notes in Pure and Applied Mathematics},
  volume={76},
  pages={113--150},
  publisher={Dekker},
  address={New York},
  year={1982}
}

@article{jayme18,
title = {The convexity of induced paths of order three and applications: Complexity aspects},
journal = {Discrete Applied Mathematics},
volume = {237},
pages = {33-42},
year = {2018},
author = {Rafael T. Araújo and Rudini M. Sampaio and Vinícius F. {dos Santos} and Jayme L. Szwarcfiter},
}

@article{dahlhaus94,
author = {Dahlhaus, E. and Johnson, D. S. and Papadimitriou, C. H. and Seymour, P. D. and Yannakakis, M.},
title = {The Complexity of Multiterminal Cuts},
journal = {SIAM Journal on Computing},
volume = {23},
number = {4},
pages = {864-894},
year = {1994},
}

@article{Harary1,
title = "Convexity in graphs",
author = "F. Harary and J. Nieminen",
journal = "Journal of Differential Geometry",
volume = "16",
pages = "185-190",
year = "1981",
}

@article{PFALTZ1,
title = "Convexity in directed graphs",
author = "J. L. Pfaltz",
journal = "Journal of Combinatorial Theory, Series B",
volume = "10",
number = "2",
pages = "143-162",
year = "1971",
address ={[{\textit{S.l.}]}},
}

@article{AMMP2024arXiv,
title = {On the hull and interval numbers of oriented graphs},
journal = {Discrete Applied Mathematics},
volume = {387},
pages = {76-91},
year = {2026},
author = {J. Araújo and A.K. Maia and P.P. Medeiros and L. Penso},
}

@article{PARKER2008641,
title = "On two-path convexity in multipartite tournaments",
journal = "European Journal of Combinatorics",
volume = "29",
number = "3",
pages = "641-651",
year = "2008",
author = "Darren B. Parker and Randy F. Westhoff and Marty J. Wolf",
}

@article{FHW1980,
title = {The directed subgraph homeomorphism problem},
journal = {Theoretical Computer Science},
volume = {10},
number = {2},
pages = {111-121},
year = {1980},
issn = {0304-3975},
doi = {https://doi.org/10.1016/0304-3975(80)90009-2},
url = {https://www.sciencedirect.com/science/article/pii/0304397580900092},
author = {Steven Fortune and John Hopcroft and James Wyllie}
}

@article{AA2021,
title = {Hull and geodetic numbers for some classes of oriented graphs},
journal = {Discrete Applied Mathematics},
year = {2022},
doi = {10.1016/j.dam.2021.03.016},
author = {J. Araujo and P.S.M. Arraes},
address = {[\textit{S.l.}]},
volume = {323},
pages = {14-27},
}

@article{CDS09,
title = "On the Convexity of Paths of Length Two in Undirected Graphs",
journal = "Electronic Notes in Discrete Mathematics",
volume = "32",
pages = "11 - 18",
year = "2009",
author = "Carmen C. Centeno and Mitre C. Dourado and Jayme L. Szwarcfiter",
}

@article{Dankelman2003,
author = {Gary Chartrand and Peter Dankelmann and Michelle Schultz and Henda Swart},
year = {2003},
title = {Twin Domination in Digraphs.},
volume = {67},
journal = {Ars Combinatoria},
address ={[{\textit{S.l.}]}},
}

@article{Arumugam2013,
author = {S. Arumugam and Karam Ebadi and Sathikala Ls},
year = {2013},
pages = {},
title = {Twin domination and twin irredundance in digraphs},
volume = {7},
journal = {Applicable Analysis and Discrete Mathematics},
address = {[\textit{S.l.}]}, 
}

@article{PWW2009,
  author    = {Darren B. Parker and
               Randy F. Westhoff and
               Marty J. Wolf},
  title     = {Convex independence and the structure of clone-free multipartite tournaments},
  journal   = {Discuss. Math. Graph Theory},
  volume    = {29},
  number    = {1},
  pages     = {51--69},
  year      = {2009},
  url       = {https://doi.org/10.7151/dmgt.1432},
  doi       = {10.7151/dmgt.1432},
  bibsource = {dblp computer science bibliography, https://dblp.org}
}

@article{PWW2006,
  author    = {Darren B. Parker and
               Randy F. Westhoff and
               Marty J. Wolf},
  title     = {Two-path convexity in clone-free regular multipartite tournaments},
  journal   = {Australas. {J} Comb.},
  volume    = {36},
  pages     = {177--196},
  year      = {2006},
  url       = {http://ajc.maths.uq.edu.au/pdf/36/ajc\_v36\_p177.pdf},
  bibsource = {dblp computer science bibliography, https://dblp.org}
}

@article{CFZ2002,
  author    = {Gary Chartrand and
               John Frederick Fink and
               Ping Zhang},
  title     = {Convexity in oriented graphs},
  journal   = {Discret. Appl. Math.},
  volume    = {116},
  number    = {1-2},
  pages     = {115--126},
  year      = {2002},
  url       = {https://doi.org/10.1016/S0166-218X(00)00382-6},
  doi       = {10.1016/S0166-218X(00)00382-6},
  bibsource = {dblp computer science bibliography, https://dblp.org}
}

@article{HW1996,
author = {Haglin, David J. and Wolf, Marty J.},
title = {On Convex Subsets in Tournaments},
journal = {SIAM Journal on Discrete Mathematics},
volume = {9},
number = {1},
pages = {63-70},
year = {1996},
doi = {10.1137/S0895480193251234},

URL = { 
        https://doi.org/10.1137/S0895480193251234
    
},
eprint = { 
        https://doi.org/10.1137/S0895480193251234
    
}
}

@article{Kante+2017,
  author = {M. M. Kante and R. M. Sampaio and V. F. {dos Santos} and  Jayme L. Szwarcfiter},
  title = {On the geodetic rank of a graph},
  year = {2017},
  journal = {Journal of Combinatorics},
  volume = {8},
  number = {2},
  pages = {323--340}
}

@article{Delire1984,
  author = {Delire, Jean-Michel},
  title = {Graphs with high Radon number},
  year = {1984},
  url = {https://www.persee.fr/doc/barb_0001-4141_1984_num_70_1_72292},
  journal = {Bulletins de l'Académie Royale de Belgique},
  volume = {70},
  number = {1},
  doi = {10.3406/barb.1984.72292},
  pages = {14--24}
}

@article{THNOAOG,
title = "The hull number of an oriented graph",
journal = "International Journal of Mathematics and Mathematical Sciences",
volume = "2003",
number = "36",
pages = "2265-2275",
year = "2003",
author = "Gary Chartrand and John Frederick Fink and Ping Zhang",
}

@article{TGNOAOG,
title = "The Geodetic Number of an Oriented Graph",
journal = "European Journal of Combinatorics",
volume = "21",
number = "2",
pages = "181 - 189",
year = "2000",
issn = "0195-6698",
author = "Gary Chartrand and Ping Zhang"
}

@article{TGNOGAD,
author="LU, Chang-hong",
title="The geodetic numbers of graphs and digraphs",
journal="Science in China Series A: Mathematics",
year="2007",
month="Aug",
day="01",
volume="50",
number="8",
pages="1163--1172",
issn="1862-2763",
}

@article{THAGNOOOG,
title = "The hull and geodetic numbers of orientations of graphs",
journal = "Discrete Mathematics",
volume = "309",
number = "8",
pages = "2134 - 2139",
year = "2009",
issn = "0012-365X",
author = "Jung-Ting Hung and Li-Da Tong and Hong-Tsu Wang"
}

@article{EITIST,
title = "Embedding tournaments in simple tournaments",
journal = "Discrete Mathematics",
volume = "2",
number = "4",
pages = "389 - 395",
year = "1972",
issn = "0012-365X",
author = "J. W. Moon"
}

@Article{SROST,
author={Erd{\H{o}}s, P.
and Fried, E.
and Hajnal, A.
and Milner, E. C.},
title={Some remarks on simple tournaments},
journal={Algebra Universalis},
year={1972},
month={Dec},
day={01},
volume={2},
number={1},
pages={238-245},
issn={1420-8911},
doi={10.1007/BF02945032},
url={https://doi.org/10.1007/BF02945032}
}

@article{THNOAG,
title = "The hull number of a graph",
journal = "Discrete Mathematics",
volume = "57",
number = "3",
pages = "217 - 223",
year = "1985",
issn = "0012-365X",
author = "Martin G. Everett and Stephen B. Seidman"
}

@article{barbosa-2012,
  author    = {Rommel M. Barbosa and
               Erika M. M. Coelho and
               Mitre C. Dourado and
               Dieter Rautenbach and
               Jayme L. Szwarcfiter},
  title     = {{On the Carath\'eodory Number for the Convexity of Paths of Order Three}},
  journal   = {SIAM Journal on Discrete Mathematics},
  volume    = {26},
  number    = {3},
  pages     = {929--939},
  year      = {2012},
  doi       = {10.1137/110828678}
}

@book{BG2008,
author = {Bang-Jensen, J{\o}rgen and Gutin, Gregory Z.},
title = {Digraphs: Theory, Algorithms and Applications},
year = {2008},
isbn = {1848009976},
publisher = {Springer Publishing Company, Incorporated},
edition = {2nd}
}

@article{CIT,
title = "Convexity in tournaments",
journal = "Bull. Société Royale des Sciences de Liège",
volume = "45",
number = "",
pages = "570 - 586",
year = "1976",
author = "J. C. Varlet"
}

@article{OCGAHNIG,
title = "Orientable convexity, geodetic and hull numbers in graphs",
journal = "Discrete Applied Mathematics",
volume = "148",
number = "3",
pages = "256 - 262",
year = "2005",
author = "Alastair Farrugia"
}

@article{DOURADO2017134,
title = {Inapproximability results and bounds for the Helly and Radon numbers of a graph},
journal = {Discrete Applied Mathematics},
volume = {232},
pages = {134-141},
year = {2017},
issn = {0166-218X},
doi = {https://doi.org/10.1016/j.dam.2017.07.032},
url = {https://www.sciencedirect.com/science/article/pii/S0166218X17303542},
author = {Mitre C. Dourado and Aline R. {da Silva}}
}
\end{document}